\documentclass[conference,compsoc]{IEEEtran}

\usepackage[table]{xcolor}
\newcolumntype{G}{@{\hspace{4pt}}!{\color{gray!45}\vrule width 0.5pt}@{\hspace{4pt}}}
\usepackage{booktabs,makecell,adjustbox}

\usepackage{algorithm}
\usepackage{algpseudocode}
\usepackage{amssymb}
\usepackage{pifont}
\usepackage[colorlinks=true, linkcolor=blue, urlcolor=blue]{hyperref}
\usepackage[most]{tcolorbox}
\usepackage{array}
\usepackage{makecell}
\usepackage{listings}
\usepackage{xcolor}
\usepackage{graphicx}
\usepackage{multirow}
\usepackage{subcaption}
\usepackage{booktabs}
\usepackage{enumitem}
\usepackage{wasysym}
\usepackage{courier}
\usepackage{xurl}
\usepackage[defaultlines=3,all]{nowidow}
\usepackage[available,functional, reproduced]{ieeebadges}

\newcommand\copyrighttext{
  \footnotesize \textcopyright 2025 IEEE. Personal use of this material is permitted. Permission from IEEE must be obtained for all other uses, in any current or future media, including reprinting/republishing this material for advertising or promotional purposes, creating new collective works, for resale or redistribution to servers or lists, or reuse of any copyrighted component of this work in other works.
}
\newcommand\copyrightnotice{
  \begin{tikzpicture}[remember picture,overlay]
    \node[anchor=south,yshift=10pt] at (current page.south)
    {\fbox{\parbox{\dimexpr\textwidth-\fboxsep-\fboxrule\relax}{\copyrighttext}}};
  \end{tikzpicture}
}

\definecolor{bugs}{HTML}{005899}
\definecolor{unsafe}{HTML}{B22222}
\definecolor{urapi}{HTML}{2E7D32}
\definecolor{lines}{HTML}{9C6B00}

\newcolumntype{B}{>{\color{bugs}}c}
\newcolumntype{S}{>{\color{unsafe}}c}
\newcolumntype{U}{>{\color{urapi}}c}
\newcolumntype{L}{>{\color{lines}}c}

\lstdefinelanguage{Rust}{
  keywords=[1]{fn, let, mut, unsafe, use, pub, impl, struct, enum, mod, as, const, static, trait, unwrap, where, type, extern, crate, macro_use},
  keywordstyle=[1]\color{blue}\bfseries,
  keywords=[2]{u32, i32, u64, i64, usize, isize, bool, char, str, Self},
  keywordstyle=[2]\color{red}\bfseries,
  keywords=[3]{if, else, match, while, for, loop, break, continue, return},
  keywordstyle=[3]\color{orange}\bfseries,
  keywords=[4]{from_bytes, section_header_nth, vec, content,
                rdiff, std, io, read, diff_and_update, try,
                set_len, as_mut_slice, through, push_str,
                from, though, write, ptr, sort, len, is_empty,
                merge_sort, merge,  do_smth, direct_call_1,
                direct_call_2, next, cmp, partial_cmp,
                my_iter, main, empty, new, foo, describe,
                format, _to_u8, _to_usize, _to_str, mem,
                panic, afl, fuzz_nohook, zero, transmute,
                as_mut_ptr, as_mut, add, copy, drop
                remove_col, init, toodee, remove_col,
                copy_nonoverlapping,  from_raw_parts_mut,
                with_capacity, count, _custom_fn0,
                get_fuzzer_byte, get_fuzzer_bool, push,
                drain_filter, from_vec, insert_row,
                parse, process, serialize, unsafe_ser,
                safe_ser, unsafe_read_byte, insert, remove, bar, baz, describe_unchecked,
                set_global_data, desc, u_desc, nonnull_raw_slice,
                capacity, push_back, shrink_to_fit, reserve_exact, swap,
                pop_back, _to_usize, _to_string,
                do_safe_ser, do_unsafe_ser, unsafe_read
                },
  keywordstyle=[4]\color{violet}\bfseries,
  sensitive=true,
  ndkeywords={Elf, ElfFile, BlockHashes,
                Read, Result, MyRead, Window,
                Ord, String, Debug, Fn, FnOnce,
                FnMut, MyTrait, Iterator, Item,
                Self, Option, Result, Ord, Diff,
                PartialEq, PartialOrd, Ordering,
                T, I, R, F, TraitA, Shape,
                CustomTy0, TooDee, None, Some,
                SliceDeque, Vec, Serializer, Parser,
                SafeSer, UnsafeSer, StructA, Slab,
                STrait, UTrait, CustomTy1, S, U,
                CustomTy},
  ndkeywordstyle=\color{teal}\bfseries,
  comment=[l]{//},
  morecomment=[s]{/*}{*/},
  commentstyle=\color{gray!10!black}\ttfamily,
  stringstyle=\color{red}\ttfamily,
  identifierstyle=\color{green!60!black},
  morestring=[b]"
}

\ifCLASSOPTIONcompsoc
  \usepackage[nocompress]{cite}
\else
  \usepackage{cite}
\fi

\begin{document}
\date{}

\title{{\fontsize{11}{12}\selectfont \textnormal{ Accepted to IEEE Symposium on Security and Privacy 2026}} \\[1ex]\Large \bf deepSURF: Detecting Memory Safety Vulnerabilities in Rust\\Through Fuzzing LLM-Augmented Harnesses}

\author{
{\rm Georgios C. Androutsopoulos}\\
Purdue University \\
gandrout@purdue.edu \\
West Lafayette, Indiana, USA \\
\and
{\rm Antonio Bianchi}\\
Purdue University \\
antoniob@purdue.edu \\
West Lafayette, Indiana, USA \\
}
\maketitle
\copyrightnotice
\begin{abstract}
Although Rust ensures memory safety by default, it also permits the use of unsafe code, which can introduce memory safety vulnerabilities if misused. Unfortunately, existing tools for detecting memory bugs in Rust typically exhibit limited detection capabilities, inadequately handle Rust-specific types, or rely heavily on manual intervention.

To address these limitations, we present deepSURF, a tool that integrates static analysis with Large Language Model (LLM)-guided fuzzing harness generation to effectively identify memory safety vulnerabilities in Rust libraries, specifically targeting unsafe code. deepSURF introduces a novel approach for handling generics by substituting them with custom types and generating tailored implementations for the required traits, enabling the fuzzer to simulate user-defined behaviors within the fuzzed library. Additionally, deepSURF employs LLMs to augment fuzzing harnesses dynamically, facilitating exploration of complex API interactions and significantly increasing the likelihood of exposing memory safety vulnerabilities.
We evaluated deepSURF on 63 real-world Rust crates, successfully rediscovering 30 known memory safety bugs and uncovering 12 previously-unknown vulnerabilities (out of which 11 have been assigned RustSec IDs and 3 have been patched), demonstrating clear improvements over state-of-the-art tools.
\end{abstract}
\section{Introduction}

Rust's focus on prioritizing memory safety while maintaining high performance makes it a compelling competitor in the systems programming space. Unlike traditional low-level languages like C/C++, which rely on manual memory management and are prone to common memory safety bugs such as buffer overflows, Rust's unique set of safety rules eliminates these risks as early as during compilation.
This feature has enabled its adoption in mainstream operating systems~\cite{github:rustForLinux, web:rustForWin}, while its public registry of over 199k crates\footnote{In Rust, a crate is a unit of code that can be a library or a binary.} highlights its growing popularity ~\cite{web:rustRegistry}.

There are scenarios in systems programming where Rust’s safety checks can be too restrictive. To handle such cases, developers can use the \texttt{unsafe} keyword to mark code blocks or functions where the compiler’s safety checks are suspended. This explicitly signals that the enclosed code may violate Rust’s memory safety guarantees~\cite{rustBook}.

Previous studies have shown that approximately one in four Rust projects use unsafe mode~\cite{10.1145/3428204}. Developers often rely on it to implement features that are not feasible in Rust's safe mode. In other cases, unsafe Rust is used to pursue better performance or due to the complexity of writing code that the Rust compiler can verify as being safe~\cite{10.1145/3377811.3380413, mccormack2024mixedmethodsstudyimplicationsunsafe}.

Unsafe Rust undermines the language’s memory safety guarantees, introducing memory safety vulnerabilities. Microsoft and Google report that over 70\% of critical bugs stem from memory safety issues~\cite{53121, microsoft2020}, leading governmental agencies such as CISA and NSA to advocate for memory-safe programming~\cite{dod2023, cisa2023}. To date, around 27\% of Rust bugs in the RustSec Advisory Database~\cite{rustsec} are related to memory corruption, resulting from incorrect use of unsafe Rust~\cite{10646812, 10.1145/3466642}. This significant percentage has motivated research into static\cite{10.1145/3477132.3483570, 10.1109/TSE.2024.3447671, 10.1145/3385412.3386036} and dynamic~\cite{github:miri, github:aflrs, github:cargo-fuzz} analysis tools to detect such issues.

Static analyzers such as Rudra~\cite{10.1145/3477132.3483570} and Yuga~\cite{10.1109/TSE.2024.3447671} aim to identify certain bug patterns in Rust source code. Although these tools successfully detect memory safety bugs, significant human involvement is required to filter their output (with high false positive rates) and to create proof-of-concept (PoC) test cases to confirm each bug.
In contrast, dynamic analysis approaches, such as fuzzing, due to their context-sensitive analysis tend to have lower false positives and can uncover a broader range of bugs~\cite{github:cargo-fuzz, github:aflrs, 10734062}.
Although fuzzers are effective at detecting memory safety violations, their utility is often limited in the Rust ecosystem, where most code is provided as libraries rather than standalone binaries that can be directly tested by the fuzzer~\cite{10.1109/ASE51524.2021.9678813}. Consequently, before fuzzing Rust libraries, it is necessary to generate appropriate harnesses that encapsulate the library's functionality and convert the fuzzer's input into well-typed, valid data types expected by the target functions. Automated harness and test generation are well-studied in other languages: Randoop~\cite{10.1145/1297846.1297902} uses execution feedback to generate Java unit tests, and Pynguin~\cite{10.1145/3510454.3516829} extends this approach to Python.

Harness generation in Rust is tedious and time-consuming due to its complex type system and syntax. To ease this burden, several tools use static analysis to automate harness generation. 
However, despite improving test coverage, these tools struggle to detect memory safety bugs. RULF~\cite{10.1109/ASE51524.2021.9678813} and FRIES~\cite{10.1145/3650212.3680348} lack support for traits, and RuMono~\cite{10.1145/3709359} does not handle closures—both crucial for modeling buggy user-defined behavior. RPG~\cite{10.1145/3597503.3639102} targets all unsafe code, including explicitly unsafe APIs, resulting in false positives. Additionally, these tools use heuristics and API dependency analysis to generate API call sequences. While these sequences can be syntactically valid, their API calls are often not semantically related and fail to reflect realistic usage patterns. Due to these limitations, although these tools have discovered bugs in various Rust crates, none of the reported issues have involved memory safety.

More recently, the rise of Large Language Models (LLMs) for code generation has prompted their use in automating test generation~\cite{rug, 10.1145/3663529.3663801, wu2023rustgen, 10179394, proptestai} and fuzzing~\cite{10.1145/3597503.3639121}. As unit test generators, LLMs tend to focus on components they infer as important—guided largely by the user's prompt. The effectiveness of this process strongly depends on prompt quality~\cite{10.1145/3660783}, a challenge amplified in Rust due to its expressive and strict type system, which complicates the generation of valid code without precise context. To address this, researchers augment prompts with static analysis metadata and break down the generation task into smaller, focused sub-tasks. RUG~\cite{rug} follows this approach, using GPT and static analysis to generate unit tests that capture complex trait relationships. While RUG’s high-quality test code improves coverage, its evaluation has not demonstrated discovery of memory safety bugs.

In summary, the limitations of existing tools, include their insufficient approach to targeting unsafe code and relevant API call sequences, as well as limited support for Rust types such as generics requiring custom implementations and closures.

To address these limitations, we present \mbox{deepSURF}, a tool that combines static analysis and LLMs to automatically generate harnesses targeting unsafe code with the goal of uncovering memory corruption vulnerabilities. \mbox{deepSURF} uses the Rust compiler's analysis to identify APIs that can reach unsafe code and generates initial harnesses accordingly. It then employs DeepSeek-R1~\cite{deepseekai2025deepseekr1incentivizingreasoningcapability} to augment these harnesses with complex API call sequences, enabling the fuzzer to explore richer execution paths and increasing the chances of exposing memory corruption in multi-step interactions.
\mbox{deepSURF} also introduces a novel approach to handling generics by generating custom types and trait implementations. This allows the fuzzer to simulate execution of user-defined code within the library’s context, further increasing the likelihood of exposing memory bugs. For unsafe traits, where implementations must uphold safety guarantees, deepSURF prompts the LLM to decide candidate types within the library that satisfy the required bounds.

\mbox{deepSURF} employs fuzzing to test the generated harnesses, shifting the burden of discovering bug-triggering conditions from the developer to the fuzzer and eliminating manual effort. To reduce false positives, it generates harnesses that invoke safe APIs capable of reaching unsafe code and configures the fuzzer to exclude crashes unrelated to memory corruption~\cite{10734062}.

We evaluate \mbox{deepSURF} on 63 real-world Rust crates and detect 42 memory safety bugs via fuzzing. Notably, 12 of these were previously-unknown and have been disclosed to the respective library authors.
Our key contributions are summarized below.

\begin{itemize}[left=0pt]
    \item \textbf{Targeting Memory Safety Bugs.} We generate LLM-augmented harnesses that exercise semantically related API call sequences capable of reaching unsafe code. Their interaction with unsafe code can lead to memory safety violations, which our approach exposes through fuzzing.
    
    \item \textbf{Innovative Handling of Generics.} We introduce a novel method for generating harnesses for APIs that require generic type arguments, including substitution with custom types and implementation of required traits. This approach simulates the insertion and execution of potentially buggy user-defined code within the Rust library’s context.
    
    \item \textbf{Enhanced Data Type Support.} Our approach includes the generation of Rust-specific function arguments, such as closures and containers of complex or generic types, unlocking the limitations of previous work and extending the range of functions that can be fuzzed.

    \item \textbf{LLM-Guided Search Space Reduction.} We use LLMs to guide decisions in large search spaces by prioritizing potentially vulnerable API call sequences and selecting among multiple candidates for generics implementing unsafe traits and for complex types requiring instantiation.

    \item \textbf{deepSURF Tool for Automatic Detection of Memory Safety Bugs.} We implement \mbox{deepSURF}, a tool that combines LLM-augmented harness generation with fuzzing to detect memory safety bugs in Rust libraries. Compared to existing approaches, \mbox{deepSURF} shows significantly improved bug-finding capability, identifying 42 memory bugs—30 previously-known and 12 newly-discovered. We release the source code of \mbox{deepSURF} publicly~\cite{deepSURFRepo}.
\end{itemize}

\section{Background}
\label{sec:background}

\textbf{Rust: Memory-Safe by Design.}
The Rust programming language enforces strong protections against memory violations through a set of strict rules imposed by the Rust compiler (\texttt{rustc}) that developers must follow. These rules ensure memory safety by identifying potential issues either at compile time, causing the compilation to fail, or at runtime, leading to controlled program termination. In Rust, every memory object is tied to a single owner (variable). Ownership can be transferred between variables or temporarily borrowed through references, all in a controlled manner~\cite{rustBook}.
Additionally, Rust introduces the concept of lifetimes~\cite{rustBook}, which ensure that references remain valid for as long as the referenced memory object exists and is accessible.
Thanks to lifetime rules, memory bugs such as use-after-frees and double-frees are impossible in safe Rust.
Finally, Rust performs bounds checking at compile or runtime, blocking out-of-bounds memory access.

\textbf{Rust Language Features.}
In addition to its ownership and lifetime rules, Rust introduces several language features that are central to its programming model.

\emph{Traits.}
Traits in Rust are similar to interfaces in Java: they specify a set of required methods (hereafter, trait functions) that a type must implement to provide a given behavior.
For example, a type implementing the \texttt{Iterator} trait must define how to produce the next element in a sequence by implementing the \texttt{next} trait function.
Traits may also define associated types—trait-local type parameters fixed per implementation—clarifying trait's input/output relationships. Traits enable reusability and flexibility by letting code operate on any type that implements the trait.

\emph{Closures.}
Closures in Rust are anonymous functions that can capture variables from their surrounding scope. They are often passed as arguments to functions, enabling flexible and dynamic behavior (e.g., passing a closure to specify a custom condition for filtering elements of a data structure).

\emph{Panics.}
A panic in Rust represents an unrecoverable program state (e.g., out-of-bounds array access, unwrapping the \texttt{None} type or requesting more memory than is available). When a panic occurs, the Rust runtime, by default, begins unwinding the stack and terminates the program's execution in a controlled way. Panics can be triggered explicitly with the \texttt{panic!} macro to indicate violated preconditions or failed assertions. For example, APIs that allow indexing into queue-like structures often begin by asserting that the index is within bounds; otherwise, they panic. A triggered panic may not always correspond to a bug in the code, but it often serves as a deliberate guard preventing more severe failures.

\textbf{Unsafe Rust: Superpowers with High Risk.}
Systems programming often requires low-level operations, such as interfacing with native C/C++ libraries—tasks that cannot be performed using ``safe" Rust alone. To support these use cases, Rust provides unsafe Rust, which relaxes safety rules to give developers low-level control at the cost of reintroducing the risk of memory safety bugs.

Developers switch to unsafe Rust using the \texttt{unsafe} keyword to begin an unsafe code block. Functions containing unsafe blocks can be marked as either safe or unsafe. If a function is explicitly marked as unsafe, its caller is responsible for ensuring the function is not invoked with arguments that could cause memory violations. In contrast, if an unsafe code block is enclosed in a safe function, the function should act as a safe wrapper for the internal unsafe operations.
Additionally, Rust defines unsafe traits for which the compiler does not enforce safety checks. Unsafe code may rely on their implementations, and the responsibility for ensuring safety lies with the developer who implements them—whether a library author or a user~\cite{rustBook}.

\textbf{Fuzzing.}
Fuzzing is one of the most effective methods for detecting memory bugs. Fuzzers generate diverse inputs and run programs against them to uncover vulnerabilities. Greybox fuzzers such as AFL++\cite{257204} are especially efficient, using coverage feedback to explore different code paths. Their bug-finding capability can be further enhanced with sanitizers. AddressSanitizer (ASan)\cite{10.5555/2342821.2342849}, for example, detects subtle memory corruption bugs—such as small buffer overflows—that may not crash the program. However, sanitizers introduce performance overhead, which can slow down fuzzing.
In Rust, the \texttt{afl.rs}\cite{github:aflrs} crate integrates the functionality of AFL++ and automates advanced features such as CmpLog~\cite{github:aflrs-pull392} and persistent mode~\cite{github:aflrs-pull137}.

\textbf{Rust Libraries and Harness Generation.}
Rust code is organized into crates, which may be either binaries or libraries~\cite{rustBook}. Binary crates are standalone executables that must define a \texttt{main} function as the program’s entry point. Library crates, however, typically lack a \texttt{main} function and instead provide reusable functionality by exposing an API of public functions for users to call.
The Rust ecosystem primarily consists of library crates, posing challenges for automated testing tools like fuzzers~\cite{10.1109/ASE51524.2021.9678813}. While binaries can be fuzzed directly, libraries require the generation of harnesses—test cases that invoke the library’s API using input provided by the fuzzer. These harnesses are compiled into executables, allowing the fuzzer to explore the library’s functionality and uncover bugs.

\textbf{LLM-Based Harness Generation.}
Large Language Models (LLMs) have shown strong capabilities in code generation by learning from large-scale code corpora~\cite{10.5555/3666122.3667065, austin2021programsynthesislargelanguage, chen2021evaluatinglargelanguagemodels}. Their ability to synthesize semantically meaningful code has motivated security researchers to use LLMs to generate fuzz harness (sometimes referred to as fuzz drivers or fuzz targets). This approach is particularly effective when bugs are triggered by chained API sequences, as LLMs can generate interactions beyond typical usage patterns, enabling deeper testing and uncovering more intricate bugs~\cite{10.1145/3658644.3670396, 10.1145/3597926.3598067}.

The effectiveness of LLM-based harness generation heavily depends on prompt quality~\cite{ossfuzz2024llm, 10.1145/3650212.3680355}. Yet, even with sufficient documentation and examples, LLMs often struggle to generate valid harnesses for complex targets like the Linux kernel~\cite{10.1145/3663529.3663784, 10.1145/3676641.3716022} or Rust libraries~\cite{rug, wu2023rustgen}. In Rust specifically, the combination of intricate trait relationships and a highly expressive type system poses major challenges. These complexities often prevent the automatic creation of valid harnesses, as accurately reasoning about such features is better suited to traditional static analysis techniques.

\textbf{Definitions.}
To aid in understanding the rest of the paper, we propose the following terminology:
\begin{itemize}[left=0pt]
    \item \textit{Unsafe Block (\textbf{UB}):} A code segment enclosed within the \texttt{unsafe} keyword, allowing operations that bypass the language's safety checks.
    \item \textit{Unsafe Function (\textbf{UF}):} A function explicitly marked with the \texttt{unsafe} keyword in its signature. Any operation within this function can bypass Rust’s safety checks. A \textit{UF} can only be called within a \textit{UB} or another \textit{UF}. 
    \item \textit{Safe Function (\textbf{SF}):} A function that is not \textit{UF}.
    \item \textit{Unsafe Encapsulating Function (\textbf{UEF}):} A \textit{SF} that contains a \textit{UB}. \textit{UEFs} are designed to encapsulate internal unsafe code within a safe wrapper, preventing its callers from being affected by unsafe behavior.
    \item \textit{Unsafe Reaching Function (\textbf{URF}):} A \textit{SF} that is not a \textit{UEF} itself but can reach unsafe code indirectly by calling other \textit{UEFs} or \textit{URFs}.
    \item \textit{Unsafe API (\textbf{UAPI}):} A publicly accessible \textit{UF} of a Rust library.
    \item \textit{Unsafe Reaching API (\textbf{URAPI}):} A publicly accessible function of a Rust library that is either a \textit{UEF} or \textit{URF}.
    \item \textit{URAPI Coverage:} The \textit{URAPI} coverage of a set of harnesses is the ratio of \textit{URAPIs} directly called in the harnesses to the total number of \textit{URAPIs} in the library.
\end{itemize}

Listing~\ref{lst:callee-resolution} shows concrete examples of Rust functions, with each function labeled by its applicable categories.

\textbf{Validity of Rust Library Fuzzing Findings.}
The fuzzing of a Rust library harness may terminate prematurely due to runtime failures such as Rust panics, or memory errors (e.g., segmentation faults). In the context of fuzzing, we refer to these events as \emph{crashes}. Not all crashes indicate a library bug, so we classify them as follows.
\begin{itemize}[left=0pt]
    \item \textit{True positives:} Crashes caused by a developer-introduced defect in the library that can be triggered under correct (contract-respecting) use of the library. In other words, the harness follows the library according to its documentation (in terms of types, preconditions, invariants), yet erroneous or unsafe behavior occurs.
    \item \textit{False positives:} Crashes that are not caused by library bugs, including (a) intended behavior when API contracts are violated (e.g., calling an API with arguments documented to cause a panic or memory error), and (b) panics or memory errors in the harness code itself (e.g., invalid type conversions, intentional panics, or introduction of incorrect unsafe code).
\end{itemize}

In Appendix~\ref{app:fp_analysis}, we provide additional justifications and empirical data regarding this distinction.

\begin{lstlisting}[language=Rust, caption={Callee target resolution influenced by trait bounds.}, label={lst:callee-resolution}]
impl UnsafeSer for StructA {
    fn process(&self) -> Vec<u8>{ unsafe { ... } } // SF, UEF
}
impl SafeSer for StructA {
    fn process(&self) -> Vec<u8>{ ... } // SF
}
fn do_safe_ser<T: SafeSer>(dat: T) -> Vec<u8>{ // SF
    dat.process()
}
fn do_unsafe_ser<T: UnsafeSer>(dat: T) -> Vec<u8>{ // SF, URF
    dat.process()
}
pub unsafe fn unsafe_read(addr: usize) -> u8{ // UF, UAPI
    /* Unsafe operations */
}
pub fn safe_ser(sA: StructA) -> Vec<u8>{ // SF
    do_safe_ser(sA)
}
pub fn unsafe_ser(sA: StructA) -> Vec<u8>{ // SF, URF, URAPI
    do_unsafe_ser(sA)
}
\end{lstlisting}

\textbf{Memory Safety Bugs in Rust Libraries.}
The use of unsafe Rust can introduce memory safety bugs due to incorrect handling of unsafe code~\cite{10646812, 10.1145/3466642}. These bugs are similar to those found in other systems programming languages, including use-after-free, double-free, and buffer overflow violations. However, in Rust, not all memory violations constitute genuine memory bugs.

In a library crate, functions not explicitly marked as \texttt{unsafe} are expected to be safe to call under all circumstances—even if they contain internal unsafe code. If a user triggers memory corruption by interacting only with such safe functions, this constitutes a genuine memory bug: the library has failed to properly encapsulate its unsafe behavior~\cite{10.1145/3477132.3483570, rust_undefined_behavior}. In contrast, if corruption results from incorrect use of a \textit{UAPI}, the fault lies with the user for violating the safety contract~\cite{rustBook, rust_safety_contract2}.
The same principle applies to unsafe traits. When users implement unsafe traits defined by a library, they are responsible for ensuring correctness. However, if the library itself provides implementations of unsafe traits for use by its clients, it must guarantee that these implementations do not expose unsafe behavior.

Thus, identifying genuine memory safety bugs in Rust libraries requires harnesses that target \textit{URAPIs} and rely on library-defined implementations of unsafe traits. Conversely, harnesses that directly invoke \textit{UAPIs} or implement unsafe traits without adhering to their safety contracts can lead to false positives.

\section{Challenges of Fuzzing Rust Libraries}
\label{sec:motivation}
Fuzzing Rust’s ecosystem, which is primarily composed of libraries, requires generating harnesses that invoke library APIs with arguments derived from the fuzzer's input. While C functions often accept arguments constructed directly from raw bytes, Rust’s expressive type system and language-specific features, such as traits, demand more sophisticated handling.

Furthermore, although Rust confines potential memory vulnerabilities to unsafe code—unlike C, where all code is vulnerable—this introduces unique challenges in identifying and reaching that code. Unsafe code in Rust is governed by safety contracts that developers are expected to uphold. A memory violation caused by misusing a \textit{UAPI} or incorrectly implementing an unsafe trait is typically attributed to the user for violating these contracts and does not constitute a true memory bug.
Moreover, Rust existing fuzzers~\cite{cargo_fuzz, github:aflrs} classify assertion failures and panics as crashes, even when they are intentionally inserted by library developers to prevent incorrect API use~\cite{10734062}. Such crashes are often false positives and increase the amount of manual triage required to analyze the fuzzer’s findings; therefore, the fuzzer should be configured to automatically filter these crashes out.

Generating harnesses to expose memory corruption vulnerabilities in Rust requires overcoming several key challenges. Based on our analysis of real-world bugs in Rust libraries, in this section, we explain these key challenges and illustrate them with concrete examples.

\textbf{Challenge 1 (C1): Targeting Unsafe Reaching APIs and Memory Corruption Vulnerabilities.}
\label{sec:challenge_C1}
The use of unsafe code can cause memory violations. As discussed in Section~\ref{sec:background}, for a memory violation to be considered a true memory bug, it must occur when the users interact solely with the library’s safe APIs. Thus, identifying the \textit{URAPIs} of a library is essential, but challenging.

A first challenge lies in determining which functions are actually part of a library’s public API, as this often requires analyzing more than just the file where a function is defined. In Rust, a function marked with the \texttt{pub} keyword may still be inaccessible to users if it resides in a private module~\cite{rust_visibility}. Additionally, public re-exports of private modules and feature-gated code—which can alter the visible API surface across builds—introduce further complexity.

Even after identifying the exported APIs, determining which ones are \textit{URAPIs} is non-trivial. Treating every API that reaches unsafe code as a \textit{URAPI} can lead to false positives during fuzzing, especially if \textit{UAPIs} are invoked directly in the harness. Avoiding this requires a more targeted control-flow analysis of the library.
This analysis presents additional challenges. As shown in Listing~\ref{lst:callee-resolution}, both \texttt{do\_unsafe\_ser} and \texttt{do\_safe\_ser} call \texttt{process}, but the method invoked depends on the trait bound—\texttt{UnsafeSer} or \texttt{SafeSer}. Resolving the correct callee requires analyzing the caller’s type context, which can be further complicated when using dynamic dispatch~\cite{rustBook}.

Finally, beyond false positives from misuse of \textit{UAPIs}, fuzzers may also classify assertion failures and panics as crashes, even when inserted intentionally by library developers to enforce invariants and API contracts~\cite{10734062}. While some of these failures may be related to genuine bugs, their majority are either false positives or bugs that are not exploitable memory corruptions. Because our focus is on detecting memory corruption bugs while minimizing the manual effort required to analyze fuzzer crashes, we should filter these failures out. We acknowledge that this design choice may lead to false negatives, and we evaluate this aspect in Appendix~\ref{app:fp_analysis}.

To address these challenges, deepSURF performs dedicated analysis to determine unsafe code reachability, identify \textit{URAPIs}, and generate fuzz harnesses that target them. These harnesses are then tested using a specially configured fuzzer that ignores crashes caused by panics or assertions that do not indicate memory corruption. In the case of Listing~\ref{lst:callee-resolution}, deepSURF generates a harness only for the \textit{URAPI} \texttt{unsafe\_ser}, and not for the \textit{UAPI} \texttt{unsafe\_read} or the function \texttt{safe\_ser} that does not reach unsafe code.

\begin{lstlisting}[caption={\texttt{TooDee} constructors and \textit{URAPI} \texttt{insert\_row}.}, label={lst:toodee-const-insert_row}]
impl<T> TooDee<T> {
  pub fn with_capacity(cap: usize) -> TooDee<T>
  pub fn init(c: usize, r: usize, val: T) -> TooDee<T>
  pub fn from_vec(c: usize, r: usize, v: Vec<T>) -> TooDee<T>
  pub fn insert_row<I>(&mut self, ...)
}
\end{lstlisting}

\textbf{Challenge 2 (C2): Supporting Complex Types.}
\label{sec:challenge_C2}
Rust supports the definition of complex types through structs and enums. Unlike in C, where such types can often be instantiated directly from raw bytes, Rust’s visibility rules typically require using specialized functions to construct instances~\cite{rust_struct_visibility}. However, Rust does not have constructors as a built-in language construct. Instead, any library API function with the appropriate inputs and output can act as a constructor for a complex type. 

Identifying APIs that serve as constructors for a given complex type requires type analysis and type matching, tasks made non-trivial by Rust's expressive type system. Even after identifying compatible constructors, selecting which ones to include in a harness remains a challenging decision.
The selection is often heuristic-driven and constrained by the number of harnesses to generate, or whether a constructor has already been used in another harness~\cite{10.1145/3597503.3639102, 10.1109/ASE51524.2021.9678813}. Our experiments show that constructor choice significantly affects both the likelihood and speed of bug discovery. However, it is not possible to statically determine which constructor will be more effective in triggering a bug.
Listing~\ref{lst:toodee-const-insert_row} shows three constructors for the \texttt{TooDee} object from the \texttt{toodee} crate, along with the \textit{URAPI} function \texttt{insert\_row}. 
The choice of constructor has a substantial impact on the time required to trigger the double-free bug in \texttt{insert\_row}: objects created with \texttt{init} trigger the bug within seconds, whereas those created using \texttt{from\_vec} may take over five hours. Interestingly, empty objects constructed with \texttt{with\_capacity} do not trigger the bug at all.

To address this, deepSURF leverages \texttt{rustc}'s type analysis to identify and extract constructor APIs for complex type arguments. These constructor candidates are passed to the integrated LLM, which uses its semantic understanding to select a heterogeneous subset. The resulting LLM-augmented harness allows the fuzzer to dynamically choose among these constructors based on its input, prioritizing those more likely to trigger bugs.

\textbf{Challenge 3 (C3): Handling Generic Types.}
Rust supports generic data types, enabling reusability through type parameters. Generic types often have trait bounds specifying the behavior a type must implement. When an API expects a generic type, the harness must substitute it with a concrete type that satisfies the required trait bounds. This substitution is complex, as trait definitions may be scattered across the target or external libraries and some bounds may depend on supertraits~\cite{rustBook}. Such complexity requires robust static analysis to extract complete trait information.

To address this challenge, deepSURF performs trait analysis at compile time to collect all required bounds for generic arguments. During harness generation, it uses this information to generate custom types with corresponding trait implementations that substitute the generic arguments. These custom types allow the fuzzer to control traits behavior based on its input. For generics bounded by unsafe traits, deepSURF prompts the integrated LLM to identify compatible library-defined types for substitution~\cite{rustBook}, thereby avoiding violations of safety contracts due to incorrect custom implementations (see Section~\ref{sec:background}).

\begin{lstlisting}[caption={PoC for the RUSTSEC-2021-0094 BOF bug.}, label={lst:rdiff-poc}]
struct MyRead(());
impl Read for MyRead {
  // Return always the same incorrect number
  fn read(&mut self, _: &mut [u8]) -> Result<usize> {Ok(999)}
}
fn main() {
  let mut hashes = BlockHashes::empty(32);
  let diff = hashes.diff_and_update(MyRead(()));
}
\end{lstlisting}

\textbf{Challenge 4 (C4): Simulating Interaction with User-Defined Code.}
Rust libraries often allow users to define custom behavior through closures and traits. User-defined closures can be passed as API arguments to enable custom dynamic behavior. In addition, users can create custom types with specific trait implementations, allowing libraries to adapt their functionality. 

Our analysis of real-world memory safety bugs shows that library developers often fail to fully account for the range of behaviors that user-defined trait implementations or closures may exhibit—especially when interacting with unsafe code. For example, if a user-defined closure panics inside an unsafe block of the library, the Rust runtime will unwind the stack and invoke destructors for all live variables. If the unsafe code has duplicated ownership of any of these variables without proper safeguards, this can lead to a double-free bug.

For instance, an incorrect custom implementation of the \texttt{read} function from the \texttt{Read} trait can trigger a buffer overflow in the \texttt{rdiff} crate (Listing~\ref{lst:rdiff-poc}). In this case, the vulnerable \textit{URAPI} \texttt{diff\_and\_update} takes a generic argument bounded by the \texttt{Read} trait. During execution, it creates a \texttt{Window} object with vector lengths determined by the user-provided \texttt{read} function, using unsafe blocks to set these lengths without validation (Listing~\ref{lst:rdiff-window}). If the developer supplies a \texttt{Read} implementation that reports reading more bytes than the buffer can hold, the library will incorrectly allocate vectors with lengths exceeding their capacity, leading to memory corruption.

To simulate the execution of user-defined code within the library’s context, deepSURF generates custom functions that substitute trait implementations or closures, adhering to the correct syntax for each case. These functions are designed so that their behavior is driven by the fuzzer, allowing them to mimic user-defined logic that may panic or return edge-case values within the expected return type—conditions that can trigger bugs in the tested library.

\begin{lstlisting}[caption={Unsafe length setting in \texttt{Window} constructor.}, label={lst:rdiff-window}]
impl<R: Read> Window<R> {
  pub fn new(mut r: R, b_sz: usize) -> Result<Window<R>>{    
    let mut back = vec![0; b_sz];
    let size = r.read(back.as_mut_slice())?;
    unsafe { back.set_len(size); }
    // Similar unsafe length set for front
    Ok(Window {
      front,
      back,
      ...
\end{lstlisting}

\textbf{Challenge 5 (C5): Supporting Sequences.}
Although some bugs can be triggered by fuzzing a single API, others require exercising sequences of API calls, where each call sets up conditions for a more complex bug. For example, Listing~\ref{lst:simple-slab-poc} shows a PoC for a heap buffer overflow bug due to an off-by-one error in the unsafe code of \texttt{remove} of the \texttt{simple-slab} crate. Notably, this bug is only triggered after inserting at least two items into a \texttt{Slab} object before calling \texttt{remove}.

The three APIs (\texttt{with\_capacity}, \texttt{insert}, and \texttt{remove}), encapsulate unsafe code and are therefore identified as \textit{URAPIs}. However, approaches that fuzz each \textit{URAPI} in isolation fail to expose the bug described above. This example highlights that targeting a single \textit{URAPI} is often insufficient; instead, harnesses must be designed to fuzz a target API within meaningful sequences of related API calls.
For instance, as shown in the example, if we want to target a method called \texttt{remove} (removing elements from a collection), it is reasonable to first call one or multiple times the method \texttt{insert} on the same collection.

Existing methods extract fixed sequences of APIs using static analysis and dependency graphs, aiming to maximize API coverage. While these sequences are often syntactically valid, they may yield API combinations that cannot represent realistic usage scenarios. To address this limitation, deepSURF leverages LLMs to generate harnesses that embed \textit{URAPIs} within semantically relevant sequences of other API calls. For example, in the case of the \texttt{simple-slab} crate, if the LLM is provided with appropriate documentation, it can infer that \texttt{Slab} implements a list-like data structure and, accordingly, group operations such as insertion and removal within the same harness.

Although LLMs are effective at generating semantically coherent API groupings, they are less suited for exploring code paths or identifying paths with high coverage or a greater likelihood of exposing bugs~\cite{rug}. To address this, deepSURF prompts the LLM to generate harnesses in which both the length and composition of API sequences are parameterized by the fuzzer’s input. This enables the fuzzer to dynamically compose sequences—skipping, repeating, or prioritizing API calls—based on runtime feedback.

\begin{lstlisting}[language=Rust, caption={PoC for the RUSTSEC-2020-0039 BOF bug.}, label={lst:simple-slab-poc}]
struct StructA(String);
fn main() {
    let mut slab = Slab::with_capacity(2);
    slab.insert(StructA("Hi".to_string()));
    slab.insert(StructA("Bye".to_string()));
    slab.remove(0); // Crashes due to heap buffer overflow
}
\end{lstlisting}

\begin{table}[htbp]
\caption{Tools comparison against the challenges (C1-C5). Symbols indicate: full (\checkmark{}), partial (\LEFTcircle) or no (\ding{55}) support.}
\footnotesize
\renewcommand{\arraystretch}{1}
\centering
\begin{tabular}{|r|c|c|c|c|c|}
\hline
\textbf{Tool} & \textbf{C1} & \textbf{C2} & \textbf{C3} & \textbf{C4} & \textbf{C5} \\ \hline
RULF & \ding{55} & \LEFTcircle & \ding{55} & \ding{55} & \LEFTcircle\\ \hline
RPG  & \LEFTcircle & \LEFTcircle & \checkmark & \ding{55} & \LEFTcircle \\ \hline
FRIES & \ding{55} & \LEFTcircle & \ding{55} & \ding{55} & \LEFTcircle \\ \hline
RuMono & \ding{55} & \checkmark & \checkmark & \ding{55} & \LEFTcircle \\ \hline
RUG & \LEFTcircle & \checkmark & \checkmark & \ding{55} & \LEFTcircle \\ \hline
\hline
deepSURF & \checkmark & \checkmark & \checkmark & \checkmark & \checkmark \\ \hline
\end{tabular}
\label{tab:challenges}
\end{table}

\textbf{Existing Solutions.}
As shown in Table~\ref{tab:challenges}, state-of-the-art Rust fuzzing tools face key limitations in addressing the explained challenges, hindering their ability to uncover memory safety bugs.

None of the existing tools configure the fuzzer to ignore crashes unrelated to genuine bugs, leading to false positives when panics or assertion failures are triggered due to incorrect API invocation.
Only RPG and RUG employ strategies to target unsafe code, but they may also directly fuzz \textit{UAPIs} without preserving their safety contracts, which can also result in false positives. While all five tools support complex type arguments, some struggle when these types are nested within container types such as vectors. Although several approaches support generic types, none supports user-defined code simulation through custom trait implementations.

Regarding \textbf{C5}, RUG generates sequences of APIs needed to construct arguments for each target function but does not embed the target function itself within broader sequences of semantically related APIs. Other tools use static analysis to build API dependency graphs and extract sequences by maximizing API coverage, prioritizing unsafe functions, or mimicking real-world usage patterns. However, these approaches impose fixed sequences that often miss ``semantic" connections between related APIs (e.g., \texttt{insert} and \texttt{remove}) and may combine unrelated or already well-tested APIs. This limits the fuzzer’s ability to dynamically explore alternative interactions that could be more effective at exposing bugs.

In summary, as we will show in Section~\ref{sec:evaluation}, while these tools improve test coverage and have identified some bugs, they have not yet demonstrated effectiveness in uncovering memory corruption vulnerabilities.

\section{deepSURF Design}
\begin{figure}[htbp]
    \centering
    \includegraphics[width=\columnwidth, clip]{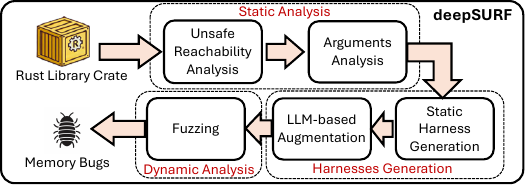}
    \caption{The workflow of deepSURF.}
    \label{fig:surf_workflow}
\end{figure}

The design of our tool is illustrated in Figure~\ref{fig:surf_workflow}. deepSURF consists of three key components: \textbf{Static Analysis}, \textbf{Harness Generation}, and \textbf{Dynamic Analysis}. The tool processes a Rust library as input and attempts to automatically detect memory safety vulnerabilities affecting it.

\subsection{Unsafe Reachability Analysis}

First of all, deepSURF needs to identify all \textit{URAPIs} in the provided Rust library, since these are the functions that can potentially trigger the execution of unsafe code. This process consists of two steps: (a) detecting all locations of unsafe code within a library and (b) finding publicly accessible safe functions that can reach these locations.

\subsubsection{Identifying Unsafe Usage and UEFs}

In a Rust library, unsafe code can be found in unsafe blocks and in unsafe functions (i.e., code blocks and functions marked by the keyword \texttt{unsafe}). deepSURF focuses specifically on unsafe code that is reachable through safe functions (\textit{SFs}), as directly calling \textit{UFs} may lead to false positives (see Section~\ref{sec:background}). Therefore, deepSURF identifies \textit{UEFs} within the library, which act as transition points from safe to unsafe code. To achieve this, deepSURF utilizes \texttt{rustc}'s unsafety checking pass to detect all occurrences of unsafe blocks. It then leverages Rust's High-Level Intermediate Representation (HIR) to identify safe functions containing these unsafe blocks, forming the set of \textit{UEFs} in the library.

\subsubsection{Detecting Public Entry Points to \textit{UEFs}}
The next step for deepSURF is to identify all safe API functions in the library that can reach the previously-extracted \textit{UEFs}. We focus on safe APIs capable of reaching \textit{UEFs} since they serve as public entry points for the harness to access unsafe code when the fuzzer provides appropriate inputs. To identify call paths between safe APIs and \textit{UEFs}, deepSURF needs to build a Control Flow Graph (CFG) for the input library through the following steps.

\textbf{Function Call Resolution.}
\label{subsubsec:rec_fn_calls}
First, deepSURF analyzes the calling relationships between functions in the library by leveraging the Mid-Level Intermediate Representation (MIR) of each function to record all caller-callee pairs. It identifies three types of function calls: (a) direct calls with a single target, (b) direct calls with multiple targets, and (c) indirect calls through function pointers. 
deepSURF does not perform pointer analysis, so type (c) is not supported. Direct calls with single target are resolved at compilation time using the context in which the call occurs. deepSURF extracts this information from \texttt{rustc} to record the corresponding caller-callee pairs. Regarding calls with multiple potential targets at compile time, which require dynamic dispatch, deepSURF handles such cases by over-approximating and recording a caller-callee pair for each possible callee.

\textbf{CFG Construction and \textit{URAPIs} Extraction.}
After collecting all caller-callee pairs, deepSURF constructs the CFG involving all functions of the library. Next, it uses the unique identifiers assigned during HIR analysis to locate the \textit{UEFs} within the CFG. Starting from each \textit{UEF}, deepSURF performs a reverse traversal of the graph using breadth-first search (BFS), moving from callees to callers until it reaches functions with no further callers. 
During this traversal, any function that is both public and safe is added to the set of \textit{URAPIs}. By the end of this process, the \textit{URAPIs} form a subset of the library’s public API: safe functions that can reach unsafe code. This set becomes the fuzzing target, addressing \textbf{C1} (see Section~\ref{sec:challenge_C1}).

\begin{table}[htbp]
\caption{Rust data types supported by deepSURF. \texttt{N} refers to an integer; \texttt{T}, \texttt{T\textsubscript{1}}, ..., \texttt{T\textsubscript{n}} refer to types from categories 1 to 8; \texttt{Tr}, \texttt{Tr\textsubscript{1}}, ..., \texttt{Tr\textsubscript{n}} refer to traits defining required behaviors; and \texttt{E} represents an error type.}

\scriptsize
\centering
\begin{tabular}{|>{\raggedright\arraybackslash}m{3.4cm}|m{\dimexpr\linewidth-3.9cm-2\tabcolsep-1.5pt}|}
\hline
\textbf{Category} & \textbf{Examples} \\ \hline
\textbf{1. Primitive Types} & Integers, Booleans, Floating-point numbers, Characters\\ \hline
\textbf{2. Standard Library Types} & Strings, Vectors (\texttt{Vec<T>}), Boxes (\texttt{Box<T>}), Options (\texttt{Option<T>}), Results (\texttt{Result<T, E>}) \\ \hline
\textbf{3. Slices} & Common (\texttt{\&[T]}) and string slices (\texttt{\&str}) \\ \hline
\textbf{4. Compound Types} & Arrays (\texttt{[T; N]}), Tuples ((\texttt{T\textsubscript{1}}, ..., \texttt{T\textsubscript{n}})) \\ \hline
\textbf{5. Complex Types} & Structs, Enums \\ \hline
\textbf{6. Reference and Pointer Types} & References (\texttt{\&T, \&mut T}), Raw Pointers (\texttt{*const T, *mut T}) \\ \hline
\textbf{7. Generic Types} & Type parameters (e.g., \texttt{G}) serve as placeholders for compatible concrete types and may require traits (\texttt{G}: \texttt{Tr\textsubscript{1}} + ... + \texttt{Tr\textsubscript{n}})\\ \hline
\textbf{8. Rust-Specific Types} & Closures, Trait-associated Types, Dynamic Trait Objects (\texttt{dyn Tr}) \\ \hline
\end{tabular}
\label{tab:surf_supported_types}
\end{table}

\subsection{Arguments Analysis}
Rust supports a diverse range of types, from primitive types such as integers to common types found in other programming languages, such as structs and generics, as well as Rust-specific types like trait types. deepSURF is designed to handle this variety to enable effective fuzzing of \textit{URAPIs} with diverse inputs, addressing the limitations of state-of-the-art tools in supporting different argument types. A complete list of the data types that deepSURF can generate using fuzzer's input is provided in Table~\ref{tab:surf_supported_types}.

\subsubsection{Complex Types Support}
Rust uses complex types such as structs and enums to define new types within a library. Since deepSURF aims to support a wide range of \textit{URAPIs}, including those requiring complex type arguments, it must identify ways to generate instances of these types to invoke the respective \textit{URAPIs} during testing.

\textbf{Instantiation of Complex Types in Rust.}
Structs group related data into a single complex type. Although structs can be initialized by directly assigning values to their public fields, this approach is discouraged as it bypasses encapsulation. Instead, struct fields in Rust are by default private, and constructor functions are used for validated initialization. Similarly, enums define a type that can represent one of several variants, each of which can optionally hold data. The variants of a public enum can be directly assigned, or constructors can be used for initialization.

\textbf{Identifying Constructors of Complex Types.}
Since constructors are not a built-in language construct in Rust, deepSURF identifies candidate constructors for structs and enums by searching for functions that meet specific criteria. We qualify a function as a candidate constructor for a complex type if it satisfies all of the following: (a) it is public, (b) none of its input arguments are of the same type as the target complex type (nor contain it as an inner type), and (c) its return type matches the target type—either directly, wrapped (e.g., in an \texttt{Option}), or as a field within a tuple.
To verify condition (a), deepSURF uses the visibility analysis provided by \texttt{rustc}. For (b) and (c), it recursively analyzes input and output types (see~\S\ref{sec:recursive_analysis}), leveraging unique type identifiers from the HIR and internal structures of \texttt{rustc}'s type checker for precise type matching. In the case of enums, deepSURF also collects and inspects their variants, analyzing any inner data types. Collectively, these steps help address \textbf{C2} (see Section~\ref{sec:challenge_C2}).

\subsubsection{Generic Types Support}
Based on our analysis of real-world memory safety bug PoCs, nearly 90\% of the involved APIs use generic arguments, most bounded by traits. Additionally, we observed that many of these bugs stem from two factors: (a) substituting these generics with user-defined types, and (b) providing faulty custom trait implementations. This highlights the prevalence of generic arguments in Rust libraries and their connection to memory bugs when misused. As deepSURF aims to generate harnesses that expose memory safety vulnerabilities, supporting generic arguments and trait bounds is essential.

\textbf{Collecting Trait Bounds.}
When handling APIs with generic type arguments, deepSURF must collect their trait bounds to ensure valid substitutions. Trait bounds define the traits a concrete type must implement to substitute the generic argument. These bounds may also include supertraits—traits a type must implement as a prerequisite to the current trait. deepSURF utilizes metadata of the trait analysis performed by \texttt{rustc} to collect the trait (and supertrait) bounds for each generic argument.

\textbf{Collecting Trait Functions and Associated Types.}
Traits in Rust define required behavior through trait functions that a type must implement to satisfy the trait. Some traits also declare associated types—types tied to the trait and used within its context. Like generic parameters, associated types can have trait bounds and must be substituted with concrete types during harness generation. Additionally, APIs may impose constraints that link associated types to generic parameters, creating dependencies between them.

To support generic type substitution required by \textbf{C3}, deepSURF collects detailed information about all relevant traits—including their trait functions and associated types—by leveraging \texttt{rustc}'s APIs. In Rust, traits with multiple functions often include default implementations, meaning only a subset must be explicitly implemented for a type. deepSURF records both the required trait functions and any default ones that can be overridden by user-defined implementations.
This information is also essential for addressing \textbf{C4}, guiding the substitution of generics and associated types with custom types and implementations of the required and optionally overridden trait functions.

\begin{algorithm}[htbp]
\scriptsize
\caption{Recursive Type Analysis}
\label{alg:analyze_arg_and_arguments}
\begin{algorithmic}[1]
\Function{Analyze\_Args}{fn}
    \ForAll{$arg \in fn.args$} \State $\textsc{Analyze\_Arg}(arg)$ \EndFor
\EndFunction
\Function{Analyze\_Arg}{arg}
    \State $atype \gets \textsc{Get\_Arg\_Type}(arg)$
    \If{$\textsc{Is\_Generic}(atype)$}
        \ForAll{$fn \in \textsc{Get\_Trait\_Fns}(arg)$}
            \State $\textsc{Mark}(fn)$; $\textsc{Analyze\_Args}(fn)$
        \EndFor
    \EndIf
    \If{$\textsc{Is\_Complex}(atype)$}
        \ForAll{$ctor \in \textsc{Get\_Constructors}(arg)$}
            \State $\textsc{Mark}(ctor)$; $\textsc{Analyze\_Args}(ctor)$
        \EndFor
    \EndIf
    \If{$\textsc{Is\_Container}(atype)$}
        \State $\textsc{Analyze\_Arg}(\textsc{Get\_Inner\_Arg}(arg))$
    \EndIf
    \State \Return $atype$
\EndFunction
\end{algorithmic}
\end{algorithm}

\subsubsection{Recursive Type Analysis}
\label{sec:recursive_analysis}
To invoke \textit{URAPIs}, the harness must construct argument types from the fuzzer’s byte stream input. While trivial for primitives, this is more complex for structs, generic types, and container types (e.g., vectors).
To ensure correct invocation of library functions, deepSURF performs recursive type analysis on function arguments during static analysis and leverages \texttt{rustc}’s type-checking system for precise type matching. For generic and complex types, deepSURF analyzes trait function or constructor arguments, respectively. For container types, it recursively analyzes inner types. The pseudocode for this recursive analysis is provided in Algorithm~\ref{alg:analyze_arg_and_arguments}.

\subsection{Static Harness Generation} 

After completing the static analysis phase, deepSURF uses the collected information to generate fuzz harnesses. It streamlines this process for each \textit{URAPI} by constructing a dependency tree that captures the relationships among the function’s arguments and their components. After building the dependency tree, deepSURF generates harnesses by performing a Depth-First Search (DFS) from the leaves up to the root (\textit{URAPI}), progressively constructing each argument along the way.

\begin{figure}[htbp]
    \centering

    \begin{subfigure}[t]{\linewidth}
        \centering
        \begin{lstlisting}[language=Rust]
pub struct Shape{ sid: u64 }
impl Shape {
  pub fn new(sid: u64) -> Self { Self { sid } }
  pub fn zero() -> Self{ Self{ sid: 0 } }
  pub fn foo<S: STrait, U: UTrait>(&self, i1: S, i2: U) {
    unsafe { ... }
  }
}
pub trait STrait{ fn desc(&self) -> String; }
pub unsafe trait UTrait{ unsafe fn u_desc(&self) -> String; }
impl STrait for Shape{
  fn desc(&self) -> String { ... }
}
unsafe impl UTrait for Shape{
  unsafe fn u_desc(&self) -> String { ... }
}
        \end{lstlisting}
        \vspace{-1em}
        \caption{Example library \texttt{foo}.}
        \label{fig:foo}
        \vspace{0.5em}
    \end{subfigure}

    \begin{subfigure}[t]{\linewidth}
        \centering
        \includegraphics[scale=0.60]{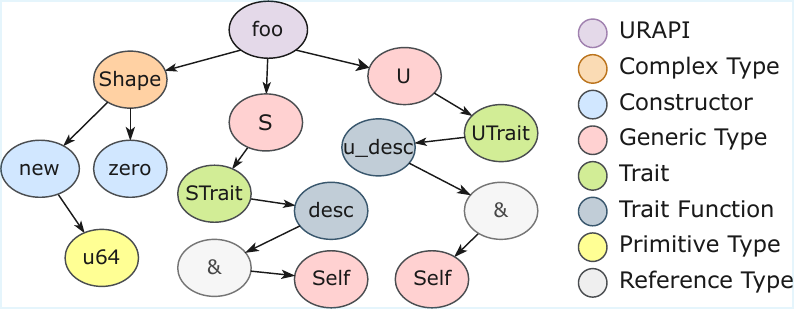}
        \caption{deepSURF’s dependency tree for \textit{URAPI} \texttt{foo}.}
        \vspace{0.5em}
        \label{fig:dependency_tree}
    \end{subfigure}

    \begin{subfigure}[t]{\linewidth}
        \centering
        \begin{lstlisting}[language=Rust]
struct CustomTy0(String);
struct CustomTy1(String);
impl STrait for CustomTy0 {
  fn desc(&self) -> String {
    if _to_u8(DATA, 0)%2 == 0{ panic!("INTENTIONAL PANIC!") }
    let mut t2 = _to_u8(DATA, 1) % 17;
    return String::from(_to_str(DATA, 2, 2 + t2 as usize));
  }
}
unsafe impl UTrait for CustomTy1 {/*Similar to STrait*/}
fn main (){
  fuzz_nohook!(|data: &[u8]| {
    if data.len() < 53 {return;}
    set_global_data(data); // DATA = data
    let t0 = foo::Shape::zero();
    let mut t7 = _to_u8(DATA, 36) % 17;
    let t8 = _to_str(DATA, 37, 37 + t7 as usize);
    let t9 = CustomTy0(String::from(t8));
    /* Similar steps with the above for building t13*/
    &t0.foo(t9, t13);
  });
}
        \end{lstlisting}
        \vspace{-1em}
        \caption{Statically generated harness for the \textit{URAPI} \texttt{foo}.}
        \label{fig:foo_harness}
    \end{subfigure}

    \caption{deepSURF’s static harness generation for \texttt{foo}.}
    \label{fig:example_combined}
\end{figure}

\textbf{Example.}
In Figure~\ref{fig:foo}, we present the example library \texttt{foo}. We focus on harness generation for the \textit{URAPI} \texttt{foo}; a simplified version of its dependency tree appears in Figure~\ref{fig:dependency_tree}. The tree’s root is the target \texttt{foo}, with child nodes for its argument types (\texttt{Shape}, \texttt{S}, \texttt{U}). These nodes connect to their corresponding constructors or trait bounds.

\textbf{Handling of Multiple Candidate Constructors.}
Dependency trees with multiple constructors per complex type are split into separate dependency trees, where each complex type is tied to a single constructor. To mitigate the exponential growth of possible trees, the number of constructors considered during this step is configurable.

\textbf{Generating Arguments.}
For primitive type nodes, deepSURF directly converts the fuzzer’s input bytes into the required types using helper functions. For generic nodes, it substitutes all occurrences of the generic parameter with custom concrete types. For complex types, it first generates the constructor arguments, which uses to invoke the constructors. Listing~\ref{fig:foo_harness} shows a harness generated for the \texttt{foo} function: the complex type \texttt{Shape} is instantiated via its \texttt{zero} constructor (line 15), while the generic parameters \texttt{S} and \texttt{U} are replaced with \texttt{CustomTy0} and \texttt{CustomTy1}, respectively—custom structs that enable trait implementation.

\textbf{Generating Custom Functions.}
To address \textbf{C4}, deepSURF generates custom implementations for all required traits. Given the trait names and trait function signatures, it synthesizes their bodies that produce return values of the expected types, using the fuzzer’s input. At this stage, it also implements unsafe traits, which temporarily introduces unsafe code into the harness—a practice we should avoid to prevent false positives (see Section~\ref{sec:background}). However, this unsafe code is removed during the LLM augmentation stage. Following a similar process, deepSURF generates custom functions to substitute closures.
As shown in Listing~\ref{fig:foo_harness}, deepSURF generates a custom implementation of the \texttt{desc} method for the trait \texttt{STrait}. To simulate user-defined behavior, the generated \texttt{desc} function uses the fuzzer’s input to either trigger a panic—potentially exposing panic-safety bugs—or return any value of the expected type.

\subsection{LLM-based Augmentation} 
\label{sec:llm-augmentation}

The harnesses generated in the previous stage represent a foundation for targeting the respective \textit{URAPIs}, but they do not fully address all the challenges we explained in Section~\ref{sec:motivation}. Specifically, they lack support for initializing complex types using multiple constructors (\textbf{C2}), fail to substitute generic parameters with unsafe trait bounds using library-defined concrete types (\textbf{C3}), and do not incorporate semantically related API call sequences that involve the targeted \textit{URAPI} (\textbf{C5}). These limitations are addressed in the current stage using LLMs.

\begin{figure}[htbp]
    \centering
    \includegraphics[width=\columnwidth]{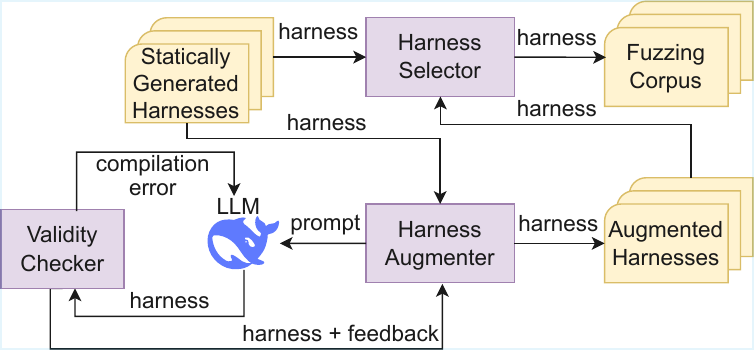}
    \caption{deepSURF's LLM harness augmentation.}
    \label{fig:augmentation_design}
\end{figure}

Figure~\ref{fig:augmentation_design} outlines the steps deepSURF performs in this stage.
In summary, the \textit{Harness Augmenter} attempts to use an LLM to improve statically generated harnesses, while the \textit{Harness Selector} optimally combines statically generated and LLM-generated harnesses.

\textbf{Prompting the Model.} 
The \textit{Harness Augmenter} selects a statically generated harness from the \textit{Statically Generated Harnesses} set, targeting a specific \textit{URAPI}, and uses it to prompt the LLM for an augmented version. When available, we prioritize fetching statically generated harnesses that compile successfully. If none exists for a given \textit{URAPI}, we fall back to non-compilable ones, which, based on our experiments, often still provide sufficient structure for the LLM to produce corrected and functional augmented versions.

Each prompt includes: (a) the fetched harness, (b) the targeted \textit{URAPI}, (c) metadata from static analysis, (d) relevant documentation, and (e) augmentation instructions. The static metadata includes available constructors for complex types and the list of \textit{URAPIs} identified in the library. Documentation—extracted via \texttt{rustdoc}—is included when input token limits allow. The instructions guide the model on how to augment the harness and outline side effects to avoid.

To fully address \textbf{C2}, we instruct the model to analyze the list of compatible constructors, select a heterogeneous subset it deems relevant, and modify the harness to support instantiating complex types using any of these constructors. The augmented harness includes a switch statement driven by the fuzzer's input to dynamically select among them.

As a final step in addressing \textbf{C3}, we prompt the model to replace custom types used for generic parameters bounded by unsafe traits with compatible, library-defined types that already implement the required traits. Importantly, we direct the model to preserve custom types wherever possible, removing only those associated with unsafe traits. This is crucial, as custom types are often paired with valuable custom trait implementations that simulate user-defined code execution (\textbf{C4}), so we aim to preserve as many of them as possible—removing only those associated with unsafe traits.

Finally, to address \textbf{C5}, we instruct the model to identify a set of semantically related APIs compatible with the targeted \textit{URAPI}, enabling the construction of meaningful API call sequences. We also request that the generated harnesses use the fuzzer's input to dynamically determine the ordering of these sequences rather than relying on fixed orderings.

\textbf{Validating Augmented Harnesses.} 
After receiving the augmented harness (response) from the LLM, the \textit{Validity Checker} uses the Rust compiler as an oracle to determine whether the harness compiles. If it does, the harness is added to the \textit{Augmented Harnesses} set, and the \textit{Harness Augmenter} is notified of the successful augmentation. In this case, the feedback to the \textit{Harness Augmenter} also includes a list of all \textit{URAPIs} invoked in the augmented harness. This list is then used to decide future harness selection for augmentation.

However, the harness may fail to compile due to errors introduced during augmentation or inherited from the original statically generated version. In such cases, the \textit{Validity Checker} retries augmentation by updating the prompt: it replaces the previous harness with the latest version and appends the relevant compiler error message before re-prompting the model. This process continues for a configurable number of tries. If all tries fail, the failure is recorded, no augmented harness is stored, and the \textit{Harness Augmenter} proceeds to the next \textit{URAPI} awaiting augmentation.

\textbf{Harnesses Selection for LLM Augmentation and Fuzzing Corpus Generation.}
deepSURF's goal is to create a corpus of fuzzing harnesses that targets unsafe code (\textbf{C1}) and captures diverse and complex behaviors within the fuzzed crate. However, generating too many harnesses risks dispersing the fuzzer's efforts and reducing its effectiveness. Furthermore, statically generated and LLM-generated harnesses often reach different code paths in the fuzzed code; thus, an appropriate combination of LLM-generated and statically generated harnesses should be included in the final \textit{Fuzzing Corpus}.
For these reasons, we apply specific \textit{harness selection policies} both in the \textit{Harness Augmenter} and in the \textit{Harness Selector} to choose which statically generated harnesses to augment and which harnesses to include in the final \textit{Fuzzing Corpus}.

Specifically, following the \textit{Harness Augmenter}'s \textit{harness selection policy} we skip augmenting a statically generated harness for \textit{URAPI A} if another already augmented harness invokes \textit{URAPI A}. This arises when the LLM augments a harness for \textit{URAPI B} by including a call to \textit{URAPI A} (e.g., when addressing \textbf{C2} or \textbf{C5}).
However, statically generated harnesses containing custom implementations are always augmented. This decision is based on our observation that such implementations, which simulate user-defined code (\textbf{C4}), cannot be reliably generated by the LLM alone. In contrast, when a statically generated harness with these implementations is provided to the LLM, the LLM can often successfully augment it while preserving its custom logic.

Regarding the \textit{Harness Selector}, we apply the following \textit{harness selection policy}.
First, we include in the final \textit{Fuzzing Corpus} all compilable LLM-augmented harnesses, as these address all \textbf{C1}–\textbf{C5}.
For \textit{URAPIs} whose augmentation fails after the maximum number of attempts, we check the set of \textit{Statically Generated Harnesses} for compilable options and add up to four harnesses if available.
Additionally, for each \textit{URAPI} involving custom functionality, we include up to four compilable statically generated harnesses, regardless of whether augmentation succeeded (\textbf{C4}).

In Section~\ref{sec:evaluation} and Appendix~\ref{app:incl_static_harnesses} we evaluate alternative harness augmentation and \textit{harness selection policies}, and we provide further justification for the chosen policies.

\subsection{Fuzzing for Memory Bugs.} 
\label{sec:fuzzing_for_mb}
deepSURF dynamically tests the generated harnesses using AFL++, delegating the task of finding bug-triggering conditions to the fuzzer. While effective at detecting memory bugs in Rust, fuzzing can produce false positives—either from panics or assertion failures intentionally inserted by library authors to enforce invariants and prevent incorrect API usage, or from errors in the harness itself (e.g., improper type conversions or incorrect use of unsafe code)~\cite{10734062, rustBook}. To prevent such cases from being treated as crashes, deepSURF modifies \texttt{afl.rs} by removing hooks that translate all panics to crashes and uses the \texttt{fuzz\_nohook!} macro as the fuzzer’s entry point. We acknowledge that this may introduce false negatives—e.g., missed logical bugs that do not lead to memory corruption—but we accept this trade-off to reduce the manual triage required to analyze our tool’s findings. We empirically evaluate this trade-off in Appendix~\ref{app:fp_analysis}.

Additionally, ASan\cite{10.5555/2342821.2342849} is enabled to detect non-crashing memory violations, and its output is used to automatically detect memory corruptions and exclude crashes caused by large memory allocations. These measures ensure that fuzzing focuses exclusively on memory corruption vulnerabilities, fully addressing \textbf{C1}.
\section{Implementation}

We implemented deepSURF in Rust, totaling approximately 13.1k lines of code. Static analysis extends the \texttt{rustc 1.81.0-dev} compiler, leveraging its query system and metadata while avoiding limitations of tools like \texttt{rustdoc}~\cite{rustc_ref, 10.1145/3709359}. Using a recent compiler ensures compatibility with modern libraries and fuzzing plugins~\cite{github:aflrs-pull392}. Below, we detail key implementation steps of our tool.

\textbf{CFG Construction and Callee Resolution.}
To perform unsafe reachability analysis, we construct a control flow graph (CFG) of all relevant library functions by emitting their MIR and iterating through basic blocks. Within these blocks, we identify \texttt{terminators}, which represent the exit points of a basic block~\cite{rustc_ref}. We specifically focus on terminators corresponding to exits caused by calls to functions. By reusing the existing MIR inliner logic, we resolve these calls, classify their types (see \S\ref{subsubsec:rec_fn_calls}), and identify locations that allow user-defined functions.

\textbf{Name Resolution and Visibility.}
The Rust compiler assigns unique identifiers to library items like functions, traits, and types during HIR analysis~\cite{rustc_ref}. However, these identifiers are internal to \texttt{rustc} and cannot be accessed in user code. deepSURF’s harnesses require access to these items to test the library, so we resolve these identifiers into publicly accessible names. Rust’s privacy rules and re-exports complicate this process. deepSURF overcomes these challenges by using metadata from \texttt{rustc}’s privacy pass~\cite{rustc_ref} to derive absolute paths for the required items.

\textbf{Data Type Analysis.}
To analyze and distinguish between data types in a Rust library, deepSURF uses \texttt{ty::Ty}, an internal \texttt{rustc} type~\cite{rustc_ref}. Its precise type information enables deepSURF to perform essential tasks, such as identifying candidate constructors and analyzing the arguments of closure types.

\textbf{Trait Information Gathering.}
Performing static analysis within \texttt{rustc} proved essential for the step of trait information collection. Specifically, by invoking \texttt{rustc}'s internal APIs, such as \texttt{predicates\_of} and \texttt{associated\_items}, deepSURF precisely gathers all the traits a generic parameter must implement, along with the trait items, including associated types and trait functions.

\textbf{Ownership Handling.}
deepSURF avoids lifetime and ownership violations through specific design choices. During static harness generation, we use a ``main" function as the entry point for the fuzzer, in which all \textit{URAPI} arguments are local variables with lifetimes scoped to main. To preserve ownership, we avoid reusing moved values by constructing each argument from distinct segments of the fuzzer’s input. Also, we ensure that LLM-augmented harnesses pass compiler checks, including lifetime validation.

\textbf{LLM Integration.}
deepSURF integrates DeepSeek-R1~\cite{deepseekai2025deepseekr1incentivizingreasoningcapability}, released in January 2025, due to its strong reasoning capabilities, cost-effectiveness, and open-source availability, which enables reproducibility of results. We specifically use version R1 (671 billion parameters, 164K-token context window), which offers a practical balance between performance, cost, and accessibility for research. If desired, deepSURF is configurable to use different LLMs. Moreover, library documentation is provided to the model using \texttt{rustdoc-md}~\cite{rustdoc-md}.

\textbf{Memory Safety of the Generated Harnesses.}
All harnesses generated by deepSURF and included in the \textit{Fuzzing Corpus} contain only safe code, enforced by the \texttt{\#![forbid(unsafe\_code)]} directive at the beginning of each harness. This prevents false positives during fuzzing that can arise from direct calls to \textit{UAPIs} that break their safety contracts, as well as including unsafe code in the harness itself. Furthermore, we verify that the use of global data complies with Rust’s safety guarantees.

\section{Evaluation}
\label{sec:evaluation}

In this section, we evaluate deepSURF by addressing three research questions and presenting two case studies of real-world memory corruption bugs detected by our tool. The research questions are as follows:

\hangindent=2.4em
\hangafter=1
\noindent \textbf{RQ1:} Can deepSURF automatically generate harnesses that uncover memory safety bugs through fuzzing?

\hangindent=2.4em
\hangafter=1
\noindent \textbf{RQ2:} How does deepSURF compare to state-of-the-art Rust fuzzing tools in detecting memory safety bugs?

\hangindent=2.4em
\hangafter=1
\noindent \textbf{RQ3:} How do the key components of deepSURF contribute to its bug-finding capability and unsafe code coverage?

\subsection{Experimental Setup}
\label{sec:setup}
To evaluate deepSURF’s effectiveness in detecting memory bugs and to compare it against state-of-the-art approaches, we construct a dataset, comprising 63 crates drawn from ERA\textsc{san}~\cite{10646812}, R\textsc{ust}S\textsc{an}~\cite{298084}, RUG~\cite{rug}, and CrabTree~\cite{10.1145/3689733} datasets.
ERA\textsc{san} and R\textsc{ust}S\textsc{an} are Rust-specific sanitizers whose datasets include crates with known real-world memory safety issues. From these datasets, we select 44 crates, excluding those that (a) for which we identified no \textit{URAPI}s, (b) are duplicated across the two datasets, (c) depend on external software, or (d) would incur prohibitive LLM-generation cost.
The remaining 19 crates are sourced from the datasets of the Rust testing tools CrabTree and RUG, which include popular, widely downloaded crates from \texttt{crates.io}.
Constructing our dataset using a sufficiently large number of crates used in prior work—including many with well-documented memory bugs across diverse libraries—provides a heterogeneous and unbiased foundation for evaluating deepSURF.

We ran our experiments on a machine equipped with AMD EPYC 7B13 CPUs (112 cores, 2.2GHz) and 224GB of memory, running Ubuntu 24.04.
In all experiments, the default deepSURF configuration uses DeepSeek-R1 with its default settings (temperature and top-p = 1.0) and sets the maximum number of prompt retry attempts to six. For static harness generation, we considered up to four constructors per complex type, we selected dependency trees using seeded random sampling and attempted to compile up to 50. We evaluate the use of different LLMs in \textbf{RQ3}.

\begin{table}[ht!]
\caption{New memory safety bugs found by deepSURF. The first column lists the RustSec ID when available. The remaining columns report the affected crate, the vulnerable function or trait, and the bug type.}
\centering
\scriptsize
\setlength{\tabcolsep}{0.6pt}
\renewcommand{\arraystretch}{1.2}
\begin{tabular}{|c|c|c|c|}
\hline
\textbf{RUSTSEC ID} & \textbf{Crate} & \textbf{Vulnerable Function/Trait} & \textbf{Bug Type} \\ \hline
RUSTSEC-2025-0062 & toodee & toodee::DrainCol::drop & BOF \\ \hline
RUSTSEC-2025-0044 & slice\_deque & SliceDeque::insert & DF \\ \hline
RUSTSEC-2025-0044 & slice\_deque & slice\_deque::IntoIter::clone & DF \\ \hline
RUSTSEC-2025-0044 & slice\_deque & SliceDeque::extend\_from\_slice & DF \\ \hline
RUSTSEC-2025-0044 & slice\_deque & SliceDeque::shrink\_to\_fit & DF \\ \hline
- & ordnung & CompactVec::remove & UAF \\ \hline
RUSTSEC-2025-0054 & array-queue & ArrayQueue::push\_front & MEMCRP \\ \hline
RUSTSEC-2025-0053 & arenavec & arenavec::common::AllocHandle & MEMCRP \\ \hline
RUSTSEC-2025-0053 & arenavec & arenavec::common::allocate\_inner & BOF \\ \hline
RUSTSEC-2025-0053 & arenavec & arenavec::common::SliceVec::split\_off & DF \\ \hline
RUSTSEC-2025-0049 & scratchpad & scratchpad::Tracking & BOF \\ \hline
RUSTSEC-2025-0050 & id-map & IdMap::from\_iter & MEMCRP \\ \hline
\end{tabular}
\label{table:deepSURF_new_bugs}
\end{table}

\subsection{Evaluation Results}

\textbf{RQ1: deepSURF's Bug-Finding Capability.}
In this experiment, we evaluate deepSURF on our dataset to determine its ability to automatically generate harnesses that uncover existing or new memory safety bugs. We fuzz each of the generated harnesses for 24 hours using two AFL++ threads working together on the same target: one with ASan and the other with CmpLog~\cite{257204}, as suggested by~\cite{aflplusplus2023fuzzing}.

deepSURF generates harnesses that detect 42 memory bugs by fuzzing (see~\cite{deepSURFRepo} for the full bug list). These include double-free (DF), buffer overflow (BOF), use-after-free (UAF), and other memory corruption violations (\mbox{MEMCRP}), such as arbitrary memory accesses and dropping of uninitialized memory. Out of the 42 found memory corruption vulnerabilities, 30 are previously-known which deepSURF detects automatically without any human involvement. In addition, deepSURF discovers 12 new memory bugs in seven crates presented in Table~\ref{table:deepSURF_new_bugs}. We have disclosed all the newly-discovered bugs to their affected developers. After our reporting, three bugs have been patched and 11 have been assigned RustSec IDs.

Many of the memory bugs detected by deepSURF involve intricate scenarios and advanced Rust features that pose challenges for automated tools. For example, triggering 21 bugs requires custom function implementations that panic or return unexpected values. Others, such as those in \texttt{slice-deque}\footnote{The crate \texttt{slice-deque} has been unmaintained since 2020, but the bugs we found also affect its maintained fork: \texttt{slice-ring-buffer}.} and \texttt{simple-slab}, are triggered only by specific API call sequences. These bug-triggering patterns are typically found only in human-written PoCs, yet deepSURF generated harnesses containing them automatically. Section~\ref{sec:casestudy2} provides more details about these harnesses.

deepSURF misses 26 reported memory bugs across 22 crates from the ERA\textsc{san} and R\textsc{ust}S\textsc{an} datasets. Several of these require language features beyond its current scope, including async programming, multithreading, or multi-step sequences that combine library APIs with internal operations (e.g., calls to \texttt{std::mem::forget}).

\begin{tcolorbox}[
    colback=white,
    colframe=black,
    arc=0mm,
    boxrule=0.1mm,
    left=0mm,
    right=0mm,
    top=0.5mm,
    bottom=0.5mm
]
\textbf{RQ1:} deepSURF identified 42 memory safety bugs (including 12 previously-unknown) by fuzzing the harnesses that it automatically generated.
\end{tcolorbox}

\textbf{RQ2: Comparison with State-of-the-Art Tools.}
We evaluated existing Rust fuzzing tools on ERA\textsc{san}'s dataset to compare their bug-finding capabilities with deepSURF. This dataset is a subset of our dataset, including 27 crates with well-documented memory safety issues.
Specifically, we ran RUG~\cite{rug}, RPG~\cite{10.1145/3597503.3639102} and RULF~\cite{10.1109/ASE51524.2021.9678813} to generate harnesses for the libraries in that dataset and fuzzed these targets to compare their findings with deepSURF. For this comparison, the other tools were tested with their original settings, while deepSURF as described in \textbf{RQ1}. We could not test RuMono~\cite{10.1145/3709359} and FRIES~\cite{10.1145/3650212.3680348}, as their code was not publicly available at the time of writing.

As shown in Table~\ref{table:comparison_deepSURF}, deepSURF detected 26 memory safety bugs in ERA\textsc{san}'s dataset, including 6 previously-unknown. None of the other tools were able to automatically expose any memory corruption bugs in the same dataset. Moreover, deepSURF achieved a high overall \textit{URAPI Coverage} (see Section~\ref{sec:background}) of 87.3\%, while RUG, RPG, and RULF achieved just 22.5\%, 4.1\%, and 3\% respectively. This significant difference highlights deepSURF’s focus on fuzzing unsafe code reachable through APIs, along with its broad support for Rust data types, enabling it to cover a substantially larger set of potentially vulnerable APIs.

RUG was the best-performing tool among the existing ones, generating harnesses of relatively high quality. While it failed to detect any memory corruption vulnerabilities when tested with its default fuzzing engine (libFuzzer~\cite{libfuzzer}), we observed that it could detect one of the bugs by just switching to AFL++ and potentially seven more after modifying its prompt. In contrast, RULF and RPG exhibit limited argument-type support, preventing them from fully addressing \textbf{C2} and \textbf{C3}. As a result, their ability to fuzz complex libraries with diverse API argument types is limited.

\begin{table}[htbp]
\caption{Bug-finding capability of deepSURF and comparison of \textit{URAPI Coverage} and number of compilable generated harnesses with other Rust fuzzing tools. — signifies cases where the tool crashes and ! denotes cases where harness generation did not finish within 24 hours.}

\centering
\scriptsize
\setlength{\tabcolsep}{0.5pt}
\renewcommand{\arraystretch}{1.2}
\begin{tabular}{|c|c|c|c|c|c|}
\hline
\multirow{2}{*}{\centering\textbf{Crate}} 
& \multicolumn{1}{c|}{\makecell{\textbf{\#Detected} \\ \textbf{Memory Bugs} \\ \textbf{(New Bugs)}}} 
& \multicolumn{4}{c|}{\textbf{URAPI Coverage (\#Harnesses)}} \\ \cline{2-6}
& \textbf{deepSURF} & \textbf{deepSURF} & \textbf{RUG} & \textbf{RPG} & \textbf{RULF} \\ \hline

algorithmica & 1 & 100\% (4) & 100\% (20) & 0\% (1) & 0\% (1) \\ \hline
arc-swap & 0 & 55.6\% (7) & 22.2\% (10) & 0\% (0) & 0\% (0) \\ \hline
tokio & 0 & 69.2\% (7) & — & 0\% (0) & 0\% (0) \\ \hline
secp256k1 & 0 & 96.5\% (92) & — & 31.4\% (56) & 20.9\% (35) \\ \hline
bumpalo & 0 & 96.4\% (48) & 64.3\% (30) & 39.3\% (29) & 35.7\% (9) \\ \hline
toodee & 4 (1) & 90\% (34) & 28.3\% (66) & 0\% (0) & 0\% (0) \\ \hline
nano\_arena & 0 & 100\% (1) & 0\% (15) & — & 0\% (0) \\ \hline
stack\_dst & 2 & 57.1\% (8) & 0\% (3) & 0\% (0) & 0\% (0) \\ \hline
slice\_deque & 5 (4) & 95.3\% (316) & 64.7\% (65) & — & 0\% (0)\\ \hline
lru & 0 & 80.8\% (18) & 0\% (0) & 0\% (0) & 0\% (0)\\ \hline
rusqlite & 0 & 84.2\% (108) & 0\% (0) & 3\% (39) & 1\% (3)\\ \hline
stackvector & 1 & 82.5\% (44) & 0\% (0) & 0\% (0) & 0\% (0) \\ \hline
insert\_many & 1 & 100\% (3) & 0\% (0) & 0\% (0) & 0\% (0) \\ \hline
smallvec-0.6.6 & 1 & 87.5\% (52) & 0\% (0) & 0\% (0) & 0\% (0) \\ \hline
smallvec-1.6.0 & 1 & 84.7\% (56) & 0\% (0) & 0\% (0) & 0\% (0) \\ \hline
futures-task-0.3.3 & 0 & 78.3\% (8) & — & 0\% (0) & 0\% (0) \\ \hline
futures-task-0.3.5 & 0 & 68.2\% (11) & — & 0\% (0) & 0\% (0) \\ \hline
simple-slab & 2 & 100\% (2) & 75\% (6) & 0\% (0) & 0\% (0) \\ \hline
ordnung & 2 (1) & 97.1\% (69) & 45.7\% (18) & — & 0\% (0) \\ \hline
cbox & 1 & 66.7\% (11) & 0\% (0) & 0\% (0) & 0\% (0) \\ \hline
string-interner & 0 & 100\% (2) & 33.3\% (8) & 33.3\% (1) & 33.3\% (1) \\ \hline
http & 0 & 86.4\% (109) & 52.5\% (173) & — & — \\ \hline
qwutils & 1 & 100\% (10) & 80\% (140) & 0\% (0) & 0\% (0) \\ \hline
endian\_trait & 1 & 100\% (9) & 100\% (59) & 0\% (0) & 0\% (0) \\ \hline
pnet\_packet & 1 & 100\% (42) & — & ! & 0\% (21) \\ \hline
rdiff & 1 & 100\% (4) & 0\% (22) & 0\% (3) & 0\% (2) \\ \hline
through & 1 & 100\% (4) & 50\% (1) & 0\% (0) & 0\% (0) \\ \hline \hline
\textbf{Total} & \textbf{26 (6)} & \textbf{87.3\% (1079)} & \textbf{22.5\%(636)} & \textbf{4.1\% (129)} & \textbf{3\% (72)} \\ \hline
\end{tabular}
\label{table:comparison_deepSURF}
\end{table}

None of the existing tools fully addresses \textbf{C4} (support for custom implementations simulating user-defined logic) or \textbf{C5} (support for generating semantically-related complex API sequences), which are critical for detecting complex memory corruption vulnerabilities.
Moreover, all three existing tools failed to properly address \textbf{C1}, since they produced false positives due to crashes unrelated to actual bugs either by directly invoking unsafe code without preserving safety invariants or due to incorrect handling of crashes caused by developer-introduced panics and assertions.
Finally, RUG and RPG failed to run on five crates, and RULF on one, due to crashes, timeouts, or errors caused by unimplemented features or implementation errors.

\begin{tcolorbox}[
    colback=white,
    colframe=black,
    arc=0mm,
    boxrule=0.1mm,
    left=0mm,
    right=0mm,
    top=0.5mm,
    bottom=0.5mm
]
\textbf{RQ2:} deepSURF outperforms state-of-the-art Rust fuzzing tools in detecting memory corruption vulnerabilities by supporting complex sequences of APIs with diverse and complex argument types, enabling it to detect 26 memory safety bugs that other tools fail to uncover.
\end{tcolorbox}

\textbf{RQ3: Effectiveness of deepSURF's Components.}
To assess how deepSURF’s core components contribute to its effectiveness in detecting memory safety bugs and fuzzing unsafe code reachable through a library’s API, we conducted an ablation study targeting four key components: (a) static analysis for static harness generation, (b) LLM-based augmentation, (c) the \textit{harness selection policies}, and (d) the choice of the LLM. We compared the full version of deepSURF against seven alternative configurations, each disabling or modifying one of these components, as summarized in Table~\ref{table:ablation_configs}.

We evaluated these configurations on three crates from the ERA\textsc{san} dataset that contain memory safety bugs requiring harnesses with diverse characteristics to be exposed. Since each configuration yields a different number of harnesses, and to ensure a fair comparison, we allocated the same total fuzzing time per configuration: 1344 CPU-hours.
Thus, the fuzzing time per harness is computed as: \texttt{\small
FuzzingTimePerHarness = 1344 / \#Harnesses}
\smallskip{}

This approach gives more fuzzing time per harness to configurations that yield fewer harnesses. We fuzz each harness using two AFL++ threads in a master-slave setup: one with ASan and the other with CmpLog. 

To measure unsafe code coverage, we count code lines within unsafe blocks or unsafe functions that are reachable through safe APIs of the evaluated crates. We exclude empty lines, comments, and feature-gated code not compiled under the default feature set. For completeness, we also report the total number of covered library code lines.

\begin{table}[htbp]
\caption{Configurations used in the ablation study. Each disables or modifies core components of deepSURF to evaluate their impact. Symbols indicate: \checkmark{} = enabled, \ding{55} = disabled, and \textemdash{} = not applicable.}
\centering
\scriptsize
\renewcommand{\arraystretch}{1.15}
\setlength{\tabcolsep}{3pt}

\begin{adjustbox}{max width=\columnwidth}
\begin{tabular}{|l|c|c|c|c|}
\hline
\multirow{2}{*}{\textbf{Configuration}} & \multicolumn{4}{c|}{\textbf{Components}}\\
\cline{2-5}
& \thead{Static Harness\\ Generation}
& \thead{LLM-based\\ Augmentation}
& \thead{Harness Selection\\ Policy}
& \thead{LLM\\ Backend}\\
\hline
deepSURF          & \checkmark & \checkmark & \makecell{Skip \textit{URAPIs} without\\ custom functionality} & DeepSeek-R1 \\ \hline
deepSURF-static   & \checkmark & \ding{55}  & \textemdash{} & \textemdash{} \\ \hline
deepSURF-llm      & \ding{55}  & \checkmark & \textemdash{} & DeepSeek-R1 \\ \hline
deepSURF-skip-all & \checkmark & \checkmark & \makecell{Skip any \textit{URAPI}} & DeepSeek-R1 \\ \hline
deepSURF-no-skip  & \checkmark & \checkmark & \makecell{Do not skip} & DeepSeek-R1 \\ \hline
deepSURF-o3  & \checkmark & \checkmark & \makecell{Skip \textit{URAPIs} without\\ custom functionality} & OpenAI o3 \\ \hline
deepSURF-sonnet4  & \checkmark & \checkmark & \makecell{Skip \textit{URAPIs} without\\ custom functionality} & Claude Sonnet 4 \\ \hline
deepSURF-gpt5  & \checkmark & \checkmark & \makecell{Skip \textit{URAPIs} without\\ custom functionality} & OpenAI GPT5 \\ \hline
\end{tabular}
\end{adjustbox}
\label{table:ablation_configs}
\end{table}

\underline{\textit{Static vs. LLM-based Analysis:}}
We distinguish the \textit{deepSURF-static} and \textit{deepSURF-llm} configurations to evaluate the individual contributions of static analysis and LLM integration to deepSURF’s ability to fuzz unsafe code and uncover memory corruption bugs.

\begin{table*}[htbp]
\caption{Ablation study results across different deepSURF configurations.}
\centering
\normalsize
\renewcommand{\arraystretch}{1.2}
\setlength{\tabcolsep}{5pt}

\begin{subtable}[t]{\textwidth}
\caption{Impact of static analysis and LLM integration.}\label{table:ablation_llm_static}
\vspace{-2mm}
\centering
\small
\setlength{\tabcolsep}{3.5pt}
\renewcommand{\arraystretch}{1.04}

\begin{adjustbox}{max width=\textwidth}
\begin{tabular}{l B S U L B S U L B S U L}
\toprule
& \multicolumn{4}{c}{\textbf{deepSURF}}
& \multicolumn{4}{c}{\textbf{deepSURF-llm}}
& \multicolumn{4}{c}{\textbf{deepSURF-static}} \\
\cmidrule(lr){2-5}\cmidrule(lr){6-9}\cmidrule(lr){10-13}
\vspace{-0.5em}
\textbf{Crate}
& \thead{Detected\\Bugs}
& \thead{Unsafe\\Coverage}
& \thead{URAPI Coverage\\(\#Harnesses)}
& \thead{\#Covered\\Code Lines}
& \thead{Detected\\Bugs}
& \thead{Unsafe\\Coverage}
& \thead{URAPI Coverage\\(\#Harnesses)}
& \thead{\#Covered\\Code Lines}
& \thead{Detected\\Bugs}
& \thead{Unsafe\\Coverage}
& \thead{URAPI Coverage\\(\#Harnesses)}
& \thead{\#Covered\\Code Lines} \\
\midrule
toodee         & 4 & 83.8\% & 90\% (34)   & 744 & 0 & 47.2\% & 35\% (9)   & 661 & 3 & 83.1\% & 81.7\% (159) & 675 \\
simple-slab    & 2 & 100\%  & 100\% (2)   & 87  & 2 & 100\%  & 100\% (12) & 89  & 0 & 50\%   & 100\% (14)   & 47  \\
smallvec & 1 & 86.8\% & 84.7\% (56) & 642 & 0 & 85.9\% & 66.7\% (44)& 620 & 0 & 0\%    & 0\% (0)      & --  \\
\rowcolor{gray!15}
\textbf{TOTAL} & 7 & 86.2\% & 87.9\% (92) & 1473& 2 & 72.2\% & 55.5\% (65)& 1370& 3 & 32.7\% & 40.7\% (173) & 722 \\
\bottomrule
\end{tabular}
\end{adjustbox}
\end{subtable}

\vspace{0.4em}

\begin{subtable}[t]{\textwidth}
\caption{Alternative Harness Selection Policies of the \textit{Harness Augmenter}.}\label{table:ablation_skip}
\vspace{-2mm}
\centering
\scriptsize
\setlength{\tabcolsep}{3.5pt}
\renewcommand{\arraystretch}{0.8}
\begin{adjustbox}{max width=\textwidth}
\begin{tabular}{l B S U L B S U L B S U L}
\toprule
& \multicolumn{4}{c}{\textbf{deepSURF}}
& \multicolumn{4}{c}{\textbf{deepSURF-skip-all}}
& \multicolumn{4}{c}{\textbf{deepSURF-no-skip}} \\
\cmidrule(lr){2-5}\cmidrule(lr){6-9}\cmidrule(lr){10-13}
\vspace{-0.5em}
\textbf{Crate}
& \multicolumn{4}{c}{\emph{\thead{As in\\Table~\ref{table:ablation_llm_static}}}}
& \thead{Detected\\Bugs}
& \thead{Unsafe\\Coverage}
& \thead{URAPI Coverage\\(\#Harnesses)}
& \thead{\#Covered\\Code Lines}
& \thead{Detected\\Bugs}
& \thead{Unsafe\\Coverage}
& \thead{URAPI Coverage\\(\#Harnesses)}
& \thead{\#Covered\\Code Lines} \\
\midrule
toodee         & \multicolumn{4}{c}{} & 2 & 91.6\% & 91.7\% (31) & 902 & 3 & 95.8\% & 95\% (49)   & 875 \\
simple-slab    & \multicolumn{4}{c}{}  & 2 & 100\%  & 100\% (3)   & 87  & 2 & 100\%  & 100\% (9)   & 88  \\
smallvec & \multicolumn{4}{c}{} & 0 & 81.5\% & 81.9\% (22) & 632 & 1 & 89.4\% & 84.7\% (49) & 652 \\
\rowcolor{gray!15}
\textbf{TOTAL} & \multicolumn{4}{c}{} & 4 & 86\%   & 87.1\% (56) & 1621& 6 & 92.2\% & 90\% (107)  & 1615 \\
\bottomrule
\end{tabular}
\end{adjustbox}
\end{subtable}

\vspace{0.4em}

\begin{subtable}[t]{\textwidth}
\caption{Impact of Alternative LLM Models.}\label{table:ablation_llms}
\vspace{-2mm}
\centering
\small
\setlength{\tabcolsep}{3.2pt}
\renewcommand{\arraystretch}{1.04}

\begin{adjustbox}{max width=\textwidth}
\begin{tabular}{l B S U L B S U L B S U L B S U L}
\toprule
& \multicolumn{4}{c}{\textbf{deepSURF}}
& \multicolumn{4}{c}{\textbf{deepSURF-o3}}
& \multicolumn{4}{c}{\textbf{deepSURF-sonnet4}}
& \multicolumn{4}{c}{\textbf{deepSURF-gpt5}} \\
\cmidrule(lr){2-5}\cmidrule(lr){6-9}\cmidrule(lr){10-13}\cmidrule(lr){14-17}
\vspace{-0.5em}
\textbf{Crate}
& \multicolumn{4}{c}{\emph{\thead{As in\\Table~\ref{table:ablation_llm_static}}}}
& \thead{Detected\\Bugs} & \thead{Unsafe\\Coverage} & \thead{URAPI\\Coverage\\(\#Harnesses)} & \thead{\#Covered\\Code Lines}
& \thead{Detected\\Bugs} & \thead{Unsafe\\Coverage} & \thead{URAPI\\Coverage\\(\#Harnesses)} & \thead{\#Covered\\Code Lines}
& \thead{Detected\\Bugs} & \thead{Unsafe\\Coverage} & \thead{URAPI\\Coverage\\(\#Harnesses)} & \thead{\#Covered\\Code Lines} \\
\midrule
toodee
& \multicolumn{4}{c}{}
& 4 & 90.8\% & 91.7\% (31) & 794
& 4 & 94.4\% & 95\% (29) & 813
& 3 & 97.2\% & 100\% (21) & 892 \\
simple-slab
& \multicolumn{4}{c}{}
& 1 & 100\% & 100\% (4) & 89
& 1 & 100\% & 100\% (3) & 88
& 2 & 100\% & 100\% (3) & 89 \\
smallvec
& \multicolumn{4}{c}{}
& 1 & 84.6\% & 91.7\% (65) & 652
& 0 & 87.7\% & 90.3\% (50) & 701
& 1 & 93\%   & 93.1\% (65) & 716 \\
\rowcolor{gray!15}
\textbf{TOTAL}
& \multicolumn{4}{c}{}
& 6 & 87.5\% & 92.1\% (100) & 1535
& 5 & 90.6\% & 92.9\% (82)  & 1602
& 6 & 94.8\% & 96.4\% (89)  & 1697 \\
\toprule
\rowcolor{gray!07}
\textbf{Cost (USD)}
  & \multicolumn{4}{c}{\$8.68}
  & \multicolumn{4}{c}{\$14.15}
  & \multicolumn{4}{c}{\$67.02}
  & \multicolumn{4}{c}{\$25.18} \\
\addlinespace[.2ex]
\bottomrule
\end{tabular}
\end{adjustbox}
\end{subtable}
\label{table:ablation_combined}

\end{table*}

In \textit{deepSURF-static}, LLM-based augmentation is disabled, and the \textit{Fuzzing Corpus} consists solely of compilable harnesses produced during the static harness generation stage. In contrast, \textit{deepSURF-llm} disables static analysis entirely—no initial harnesses are statically generated. Instead, the model is prompted to perform control-flow analysis to identify \textit{URAPIs} and synthesize harnesses from scratch. In this case, we use a similar prompting strategy to the full deepSURF setup, but instead of seeding the LLM with a statically generated harness, we provide a simple template that specifies the expected fuzzer macro structure and includes examples for converting fuzzer bytes into basic Rust types (e.g., primitives and strings). Also, no static analysis metadata is included in the prompt. The results of the comparison are summarized in Table~\ref{table:ablation_llm_static}.

Based on our results, deepSURF achieves the highest bug-finding capability (seven bugs) and the highest unsafe code coverage (86.2\%) compared to the other two configurations, demonstrating that the combination of static analysis and LLM-based augmentation outperforms approaches that rely on only one of these components.

We observed that \textit{deepSURF-llm} struggled to identify \textit{URAPIs}, particularly in the case of \texttt{toodee}, where it returned APIs that could not reach unsafe code and failed to generate harnesses for 65\% of the crate’s \textit{URAPIs}. Although it achieved high unsafe code coverage in \texttt{smallvec}, it did not trigger the bug, which required custom trait implementations—a feature supported by deepSURF through its static analysis. In contrast, for \texttt{simple-slab}, \textit{deepSURF-llm} performed comparably to deepSURF, discovering both memory safety bugs. Upon inspection, we found that \texttt{simple-slab} has a relatively simple API (e.g., lacking complex trait relationships), and its bugs depend primarily on sequences of API calls, a feature supported in both deepSURF and \textit{deepSURF-llm}.

On the other hand, \textit{deepSURF-static} detected only three bugs in \texttt{toodee} and achieved much lower unsafe code coverage than the other configurations. Its success in \texttt{toodee} is attributed to static control flow analysis and support for custom trait implementations. However, it failed to find any bugs in \texttt{simple-slab} due to the lack of sequence support. Also, it could not find the bug in \texttt{smallvec}, where all \textit{URAPIs} require generic arguments implementing the unsafe \texttt{Array} trait, feature supported only by the LLM-based augmentation. Consequently, it was unable to generate valid harnesses for \texttt{smallvec}.

\underline{\textit{Comparing Harness Policies:}}
In Section~\ref{sec:llm-augmentation}, we described the operation of the \textit{Harness Selector} and \textit{Harness Augmenter}, along with their corresponding \textit{harness selection policies}. These policies influence both the composition and size of the final \textit{Fuzzing Corpus} and, consequently, affect deepSURF's bug-finding capability. In this section, we evaluate alternative \textit{harness selection policies} for both components.

We consider two additional configurations: \textit{deepSURF-skip-all} and \textit{deepSURF-no-skip}. In \textit{deepSURF-skip-all}, the \textit{Harness Augmenter}'s \textit{harness selection policy} is modified to skip augmentation for a statically generated harness if its target \textit{URAPI} has already been invoked in a previously augmented harness—regardless of whether the current harness includes custom functionality. In contrast, \textit{deepSURF-no-skip} disables the skip logic entirely, augmenting all statically generated harnesses.
In both configurations, the \textit{Harness Selector}'s \textit{harness selection policy} still falls back to statically generated harnesses when LLM-based augmentation fails. However, unlike the default deepSURF configuration (see Section~\ref{sec:llm-augmentation}), neither configuration retrieves statically generated harnesses for \textit{URAPIs} whose arguments support custom user-defined implementations.

The results of this comparison are presented in Table~\ref{table:ablation_skip}. Among all \textit{harness selection policies}, deepSURF's default policy achieves the highest bug-finding capability. The second-best configuration is \textit{deepSURF-no-skip}, which detects six bugs. The only missed bug is a double-free in \texttt{toodee}, which deepSURF identifies using a statically generated harness with custom trait implementations. As discussed earlier, \textit{deepSURF-no-skip} does not retrieve such harnesses from the static generation stage.
Upon inspecting the \textit{Fuzzing Corpus} of \textit{deepSURF-no-skip}, we found that an LLM-augmented harness targeting the same \textit{URAPI} was present and included the necessary custom trait implementations. However, this harness was more complex, invoking multiple vulnerable \textit{URAPIs}, whereas the statically generated counterpart invoked only the specific \textit{URAPI} responsible for the missed bug. As a result, due to the limited fuzzing time and potential bug shadowing ~\cite{10188628}, the fuzzer failed to trigger that bug.

On the other hand, \textit{deepSURF-skip-all} misses two additional bugs compared to \textit{deepSURF-no-skip}. This occurs because it skips augmentation of statically generated harnesses that involve custom implementations, based solely on the fact that the corresponding \textit{URAPI} has already been invoked in another augmented harness. The issue in these cases is that the LLM is unaware of the custom implementations derived from our tool's static analysis, as they are encoded only in the statically generated harnesses. As a result, the LLM calls the \textit{URAPI} using its own heuristics, which may fail to expose the bug.

Finally, we observe that deepSURF achieves lower unsafe code coverage than \textit{deepSURF-no-skip}, yet it discovers more bugs within the same set of \textit{URAPIs}. This highlights that while high unsafe code coverage allows exploration of more potentially vulnerable regions of a Rust library, it does not necessarily result in finding more bugs. Our experiments show that bug-finding capability depends not only on the amount of unsafe code covered, but also on the context in which that code is exercised. This context is influenced by the types of arguments passed to the APIs and the specific sequences in which those APIs are invoked.

\underline{\textit{Comparing LLMs:}}
To assess the impact of the LLM backend on deepSURF’s performance, we evaluate three additional configurations: \textit{deepSURF-o3}, \textit{deepSURF-sonnet4}, and \textit{deepSURF-gpt5}. The results (see Table~\ref{table:ablation_llms}) show that deepSURF demonstrates robustness across all evaluated LLMs, detecting most of the memory safety bugs and achieving high unsafe code and \textit{URAPI} coverage. We observe that \textit{deepSURF-sonnet4} tends to propose more complex harnesses than the other configurations, which may cause augmentation to fail within the selected number of tries (see Section~\ref{sec:setup}). Overall, the default configuration that uses DeepSeek-R1 detects the most bugs while incurring the lowest cost for harness augmentation. However, we note that the other tested LLMs, achieved higher unsafe code coverage.

\begin{tcolorbox}[
    colback=white,
    colframe=black,
    arc=0mm,
    boxrule=0.1mm,
    left=0mm,
    right=0mm,
    top=0.5mm,
    bottom=0.5mm
]
\textbf{RQ3:} deepSURF performs best when combining static analysis with LLM integration, as each component contributes uniquely and synergistically to its effectiveness.
\end{tcolorbox}

\subsection{Case Studies}
In this section, we discuss two memory corruption vulnerabilities that deepSURF discovered in the crate \texttt{slice-deque}. This crate implements a double-ended queue using the \texttt{SliceDeque} type. We present two case studies: \textbf{Case Study 1} analyzes a newly discovered double-free bug, while \textbf{Case Study 2} revisits a known bug. Both case studies adhere to the APIs’ documentation and reflect realistic usage. In fact, our harnesses create \texttt{SliceDeque} objects using constructor APIs (\texttt{new}, \texttt{with\_capacity}, \texttt{from}), mutate the objects' size/capacity by calling the appropriate methods (\texttt{push\_back}, \texttt{pop\_back}, \texttt{reserve\_exact}, \texttt{shrink\_to\_fit}), and occasionally filter their elements with \texttt{drain\_filter}, matching documented usage examples.

\textbf{Case Study 1: Triggering Double-Free via Sequences of Relevant APIs.}
The \texttt{shrink\_to\_fit} \textit{URAPI} contains the unsafe code shown in Figure~\ref{lst:df-shrink_to_fit}. This function creates a smaller \texttt{new\_sdeq}, copies elements from \texttt{self} into it, and then swaps the two objects so that \texttt{self} points to the newly allocated, shrunk memory. If the inner elements are heap-allocated, both \texttt{self} and \texttt{new\_sdeq} end up pointing to the same heap memory. When \texttt{shrink\_to\_fit} returns, \texttt{new\_sdeq} goes out of scope and deallocates its memory—including the inner elements. This leaves \texttt{self} with dangling pointers to already-freed memory. As a result, when \texttt{self} is later dropped, it attempts to deallocate the same memory again, leading to a double-free bug.

\begin{figure}[htbp]
    \centering
    \begin{subfigure}[t]{0.48\textwidth}
        \begin{lstlisting}[language=Rust]
let mut new_sdeq = Self::with_capacity(self.len());
if new_sdeq.capacity() < self.capacity() {
  unsafe {
    crate::ptr::copy_nonoverlapping(
      self.as_mut_ptr(),
      new_sdeq.as_mut_ptr(),
      self.len(),
    );
    new_sdeq.elems_ =
        nonnull_raw_slice(new_sdeq.buf.ptr(), self.len());
    mem::swap(self, &mut new_sdeq);
  }
}
        \end{lstlisting}
        \vspace{-1em}
        \caption{The vulnerable \textit{URAPI} \texttt{shrink\_to\_fit} of \texttt{slice-deque}.}
        \vspace{0.4em}
        \label{lst:df-shrink_to_fit}
    \end{subfigure}
    \hfill
    \begin{subfigure}[t]{0.48\textwidth}
        \begin{lstlisting}[language=Rust]
fn main() {
  fuzz_nohook!(|data: &[u8]| {
  ...
  let constructor_sel = _to_u8(DATA) % 2;
  let mut deque = match constructor_sel {
    0 => SliceDeque::new(),
    _ => SliceDeque::with_capacity(_to_usize(DATA)),
  };
  let ops_count = _to_u8(DATA) as usize;
    for _ in 0..ops_count {
      let op = _to_u8(DATA) % 4;
      match op {
        0 => {deque.push_back(CustomTy(_to_string(DATA)));}
        1 => {deque.pop_back(); }
        2 => {deque.shrink_to_fit();}
        _ => {deque.reserve_exact(_to_usize(DATA));}
      }
  ...
}
        \end{lstlisting}
        \vspace{-1em}
        \caption{deepSURF's simplified harness that targets \texttt{shrink\_to\_fit}.}
        \label{lst:df-poc-shrink_to_fit}
    \end{subfigure}
    \caption{Triggering DF via API sequence.}
    \label{fig:df-comparison}
\end{figure}

Triggering this bug requires satisfying the condition in line 2, which ensures that \texttt{self} has greater capacity than \texttt{new\_sdeq}. Additionally, \texttt{self} must be non-empty, otherwise, no data is copied and no pointer aliasing occurs. Since \texttt{shrink\_to\_fit} does not modify the capacity or contents of \texttt{self}, other APIs must be called beforehand to establish the necessary state. For example, \texttt{reserve\_exact} increases capacity, while \texttt{push\_back} populates the queue.

By leveraging the LLM’s ability to group semantically related APIs, deepSURF generates a harness that combines all three functions. This enables the fuzzer to dynamically discover the correct call sequence and argument values, ultimately triggering a double-free bug using a harness similar to the simplified shown in Figure~\ref{lst:df-poc-shrink_to_fit}.

\textbf{Case Study 2: Triggering Double-Free via User-Defined Code Simulation.}
\label{sec:casestudy2}
The \texttt{drain\_filter(pred)} \textit{URAPI} returns an iterator (\texttt{DrainFilter}) with elements satisfying the user-defined closure \texttt{pred}. An ill-crafted \texttt{pred} can trigger a double-free bug when the iterator is consumed. This occurs if the \texttt{if}-clause (lines 7-9) executes first, followed by the \texttt{else}-clause, and a panic in \texttt{pred}, causing the \texttt{SliceDeque} destructor to free an object twice due to ownership duplication on line 12 (see Figure~\ref{lst:slice-deque-bug}). 

\begin{figure}[htbp]
    \centering
    \begin{subfigure}[t]{0.48\textwidth}
        \begin{lstlisting}[language=Rust]
fn next(&mut self) -> Option<T> {
  unsafe {
    while self.idx != self.old_len {
      let i = self.idx;
      self.idx += 1;
      ...
      if (self.pred)(&mut v[i]) {
        self.del += 1;
        return Some(ptr::read(&v[i]));
      } else if self.del > 0 {
        ...
        ptr::copy_nonoverlapping(src, dst, 1);
      }
    }
    None
  }
}
        \end{lstlisting}
        \vspace{-1em}
        \caption{The \texttt{next} function of the \texttt{DrainFilter} iterator.}
        \vspace{0.5em}
        \label{lst:slice-deque-bug}
    \end{subfigure}
    \hfill
    \begin{subfigure}[t]{0.48\textwidth}
        \begin{lstlisting}[language=Rust]
fn _custom_fn0(arg0: &mut CustomTy) -> bool {
  if get_fuzzer_byte(DATA, arg0) % 2 == 0{
    panic!("INTENTIONAL PANIC!");
  }
  return get_fuzzer_bool(DATA, arg0);
}
fn main (){
  fuzz_nohook!(|data: &[u8]| {
    ...
    let mut t1 = Vec::with_capacity(_to_usize(DATA));
    t1.push(CustomTy(_to_string(DATA)));
    ...
    let mut t135 = SliceDeque::from(&mut t1[..]);
    let mut t136 = _custom_fn0;
    let t138 = (&mut t135).drain_filter(t136).count();
  });
}
        \end{lstlisting}
        \vspace{-1em}
        \caption{deepSURF's simplified harness that targets \texttt{drain\_filter}.}
        \label{lst:slice-deque-fuzz_target}
    \end{subfigure}
    \caption{Triggering DF by simulating user-defined code.}
    \label{fig:drain-filter-comparison}
\end{figure}

This bug requires multiple conditions to hold to be triggered, and previous works have emphasized the difficulty of writing a harness for it without human intervention~\cite{10734062}. deepSURF detects this bug automatically by fuzzing the harness shown in Figure~\ref{lst:slice-deque-fuzz_target}. This harness constructs a \texttt{SliceDeque} object based on a fuzzer-driven vector (line 13). An iterator is then defined on line 15, using \texttt{\_custom\_fn0} as a substitute for the closure \texttt{pred}. This custom function utilizes the fuzzer's input to either return a random boolean or trigger a panic. This harness enables the fuzzer to uncover all conditions and patterns necessary to expose the bug.

\section{Limitations and Future Work}

deepSURF demonstrates strong capabilities in testing Rust libraries but has limitations that present opportunities for future improvement and exploration.
Currently, deepSURF uses static heuristics to decide which harnesses to fuzz, without prioritizing those that maximize coverage or expose bugs during fuzzing. In future work, we plan to incorporate fuzzer feedback to dynamically prioritize promising harnesses and promptly terminate less effective ones.
Additionally, our dataset contains 26 known memory safety bugs that deepSURF cannot detect. Upon investigation, we found no common bug-finding strategy that could address multiple of these cases, as each exhibited unique characteristics. Fine-tuning the LLM is one possible direction for future work to capture missed bug patterns in augmented harnesses. Finally, deepSURF cannot detect bugs that do not lead to memory corruption, resulting in false negatives (see Appendix~\ref{app:fp_analysis}).
\section{Related Work}

Several approaches have been proposed by researchers to identify bugs in Rust code. Static analysis tools such as Rudra~\cite{10.1145/3477132.3483570} and Yuga~\cite{10.1109/TSE.2024.3447671} scan Rust source code to identify specific bug patterns and have uncovered numerous memory safety issues~\cite{rudra_poc, yuga_repo}. 
However, these tools are limited to detecting specific bug types, suffer from a high false positive rate, and require manual effort to validate results and develop PoCs. Tools like Kani~\cite{9794041} and RustBelt~\cite{10.1145/3158154} aim to formally verify Rust code but offer limited support for memory corruption bugs during stack unwinding panics, which are cases that deepSURF can detect.

Dynamic analysis, which tests programs by executing them, has also been used in this domain. Specifically, sanitizers like ASan~\cite{10.5555/2342821.2342849} instrument code to detect memory bugs during execution but introduce performance overhead. ERA\textsc{san}~\cite{10646812},  R\textsc{ust}S\textsc{an}~\cite{298084} and L\textsc{ite}RS\textsc{an}~\cite{xia2025litersanlightweightmemorysafety} mitigate this overhead by leveraging safety checks already performed by \texttt{rustc} to avoid redundant instrumentation. However, sanitizers cannot detect bugs on their own, but they must be integrated with fuzzers or manually written test cases. Greybox fuzzers such as AFL++\cite{257204} and libFuzzer\cite{libfuzzer} use coverage feedback to discover more bugs and are available in Rust via specific crates~\cite{cargo_fuzz, github:aflrs}. FourFuzz~\cite{Paa2025} leverages AFL++ and selectively instruments Rust programs to focus fuzzing on functions that can reach unsafe code. However, since Rust code is typically distributed as libraries, it must first be harnessed.
SyRust~\cite{10.1145/3453483.3454084} synthesizes well-typed API call sequences by modeling Rust’s type system, but its scalability and lack of input mutation limit its testing power. CrabTree~\cite{10.1145/3689733} extends SyRust with support for traits, closures, and integrates fuzzing, using feedback to guide test synthesis toward sequences that improve coverage and introduce new types. Although deepSURF detects only one of the four memory safety bugs reported by CrabTree on its dataset, it does so fully automatically. By contrast, CrabTree relies heavily on developer-provided input templates that require understanding of library internals, reducing automation.

Some works have aimed to automate Rust fuzz harness generation. RULF~\cite{10.1109/ASE51524.2021.9678813} models API relationships with a dependency graph to generate harnesses. FRIES~\cite{10.1145/3650212.3680348} improves RULF’s scalability by fuzzing API sequences that reflect real-world usage. RPG~\cite{10.1145/3597503.3639102} and RuMono~\cite{10.1145/3709359} aim to maximize code coverage by supporting generics and traits, with RPG also targeting unsafe code. However, none has detected memory corruption bugs. RUG~\cite{rug} combines LLMs and fuzzing to generate high-quality unit tests but primarily focuses on improving test coverage and is not well-suited for detecting memory safety bugs. \mbox{OSS-Fuzz-Gen}~\cite{ossfuzz-gen} leverages LLMs to automatically generate fuzz harnesses. While it provides strong support for well-tested languages such as C/C++, its Rust support remains generic rather than tailored to Rust-specific features.
deepSURF overcomes these limitations by targeting key challenges that, based on our analysis of real-world Rust vulnerabilities, underlie memory violations.

Automated harness and test generation have been well explored in other languages. Randoop for Java~\cite{10.1145/1297846.1297902} and Pynguin for Python~\cite{10.1145/3510454.3516829} generate unit tests using feedback from program executions. proptest.ai~\cite{proptestai} uses LLMs to synthesize property tests for Python APIs, while Fuzz4All~\cite{10.1145/3597503.3639121} leverages LLMs to build a universal fuzzer that generates and mutates inputs for programs written in multiple languages. deepSURF builds on these ideas and brings them to the Rust ecosystem—combining static analysis, fuzzing, and LLMs—to automatically generate and test harnesses and identify memory corruption vulnerabilities, whose exploitability is well studied and can lead to costly incidents.
\section{Conclusion}

Fuzzing Rust for memory bugs is challenging due to its complex type system and the need for harness generation. In this work, we present deepSURF, a tool that automatically generates and fuzzes LLM-augmented harnesses to expose memory bugs in Rust libraries. deepSURF targets code with potential memory corruption risks, introduces novel approaches to support Rust data types and leverages LLMs to support complex sequences of semantically related APIs. Evaluated on 63 crates, deepSURF detected 42 memory corruption vulnerabilities, including 12 previously-unknown, outperforming existing tools.
\section*{Acknowledgment}
We thank the anonymous reviewers and the shepherd for their valuable comments and suggestions. This work was supported, in part, by the Eugenides Foundation-Marianthi Simou’s Legacy (Athens, Greece).
\bibliographystyle{plain}
\bibliography{main}
\appendices

\section{Harnesses Selector Policies}
\label{app:incl_static_harnesses}

When experimenting with LLM-based augmentation, we observed that, despite prompting the LLM against it, the LLM often removes statically generated custom functionality during augmentation. However, such custom logic can be crucial for exposing memory bugs during fuzzing since it allows simulation of user-defined code execution (\textbf{C4}). To avoid missing bugs that may be triggered only by statically generated harnesses, the \textit{harness selection policy} of the \textit{Harness Selector} includes up to four of the corresponding statically generated harnesses in the \textit{Fuzzing Corpus} for all \textit{URAPIs} with custom functionality.

Regarding the selection of up to four harnesses per \textit{URAPI}, we experimented with \textit{deepSURF-static} (see Section~\ref{sec:evaluation}) generating different numbers of harnesses  for the crates in ERA\textsc{san}'s dataset by selecting varying numbers of dependency trees. Specifically, we tested configurations generating up to one (1fh), two (2fh), and four (4fh) harnesses per \textit{URAPI}, and fuzzed them for six hours. Table~\ref{table:surf_on_erasan_abl} summarizes the vulnerabilities detected in each case.
Results show that 4fh achieved the best results. For this reason, we select up to four statically generated harnesses per \textit{URAPI} for inclusion in the \textit{Fuzzing Corpus} when the \textit{URAPI} involves custom functionality or when LLM-based augmentation fails.

\begin{table}[htbp]
\caption{Bug detection capability of deepSURF without LLM-based harness augmentation. Each row corresponds to a unique bug. \checkmark{} indicates detection via fuzzing, \ding{55} indicates failure to detect. * signifies a new bug discovered by deepSURF. When available, the assigned RustSec ID is provided.}
\centering
\scriptsize
\setlength{\tabcolsep}{4.5pt}
\renewcommand{\arraystretch}{1.2}
\begin{tabular}{|c|c|c|c|c|c|}
\hline
\textbf{RUSTSEC ID} & \textbf{Crate} & \textbf{Bug Type} & \textbf{1fh} & \textbf{2fh} & \textbf{4fh} \\ \hline
RUSTSEC-2021-0018 & qwutils & DF & \checkmark & \checkmark & \checkmark \\ \hline
RUSTSEC-2021-0028 & toodee & BOF & \checkmark & \checkmark & \checkmark \\ \hline
RUSTSEC-2021-0028 & toodee & DF & \ding{55} & \checkmark & \checkmark \\ \hline
RUSTSEC-2025-0062* & toodee & BOF & \ding{55} & \checkmark & \checkmark \\ \hline
RUSTSEC-2025-0044* & slice\_deque & DF & \checkmark & \checkmark & \checkmark \\ \hline
RUSTSEC-2025-0044* & slice\_deque & DF & \checkmark & \checkmark & \checkmark \\ \hline
RUSTSEC-2021-0047 & slice\_deque & DF & \ding{55} & \checkmark & \checkmark \\ \hline
RUSTSEC-2021-0094 & rdiff & BOF & \checkmark & \checkmark & \checkmark \\ \hline
RUSTSEC-2021-0053 & algorithmica & DF & \checkmark & \checkmark & \checkmark \\ \hline
RUSTSEC-2021-0049 & through & DF & \checkmark & \checkmark & \checkmark \\ \hline
RUSTSEC-2021-0039 & endian\_trait & DF & \checkmark & \checkmark & \checkmark \\ \hline
RUSTSEC-2020-0167 & pnet\_packet & BOF & \checkmark & \checkmark & \checkmark \\ \hline
RUSTSEC-2020-0005 & cbox & MEMCRP & \checkmark & \checkmark & \checkmark \\ \hline
RUSTSEC-2021-0042 & insert\_many & DF & \checkmark & \checkmark & \checkmark \\ \hline
RUSTSEC-2020-0038 & ordnung & DF & \ding{55} & \ding{55} & \checkmark \\ \hline
* & ordnung & UAF & \ding{55} & \ding{55} & \checkmark \\ \hline
\end{tabular}
\label{table:surf_on_erasan_abl}
\end{table}

\section{False Positives Analysis}
\label{app:fp_analysis}

As noted in Section~\ref{sec:background}, fuzzing Rust harnesses can yield crashes that are not genuine bugs (false positives). deepSURF targets automatic detection of memory safety bugs in Rust libraries while keeping manual triage effort low. An output dominated by false positives would undermine the tool’s utility. To mitigate this, deepSURF configures \texttt{afl.rs} to ignore panics that do not lead to memory corruption (see  Section~\ref{sec:fuzzing_for_mb}). In this section, we empirically evaluate the effectiveness of this design choice.

We introduce \textit{deepSURF-no-filter}, a configuration of deepSURF that keeps panic hooks enabled (see Section~\ref{sec:fuzzing_for_mb}). We fuzz both deepSURF and \textit{deepSURF-no-filter} using the settings and crates of our ablation study (see Section~\ref{sec:evaluation}). deepSURF produces 630 total crashes; \textit{deepSURF-no-filter} produces 5{,}764. Leveraging ASan's instrumentation which forces a crash when memory violation occurs during fuzzing, both variants auto-classify memory corruption crashes—607 for deepSURF and 292 for \textit{deepSURF-no-filter}—all true positives. These deduplicate into seven and six distinct memory safety bugs, respectively. After removing the ASan-confirmed cases, 23 crashes (deepSURF) and 5{,}472 crashes (\textit{deepSURF-no-filter}) remain for manual triage. For deepSURF, all 23 are false positives and can be dismissed within minutes of human effort. For \textit{deepSURF-no-filter}, 5{,}118 of the remaining crashes are false positives, requiring at least a day of manual work to classify.

Moreover, \textit{deepSURF-no-filter} triggers 354 crashes attributable to 12 distinct arithmetic bugs (e.g., addition overflows) in the fuzzed crates. Because \texttt{afl.rs} enables arithmetic overflow checks by default, these cases appear as panics during fuzzing. Notably, four of these arithmetic bugs occur inside \texttt{unsafe} blocks and could potentially lead to memory corruption when building in release mode, where overflow checks are typically disabled. By design, deepSURF misses these 12 bugs (therefore, they can be considered as false negatives), due to the fact that in its current version panics from arithmetic overflows are ignored. However, it could be configured to detect the ones that manifest as memory bugs by compiling the harnesses with the build setting \texttt{overflow-checks=off}.
\section{Meta-Review}

The following meta-review was prepared by the program committee for the 2026
IEEE Symposium on Security and Privacy (S\&P) as part of the review process as
detailed in the call for papers.

\subsection{Summary}
This paper presents deepSURF, a system that automatically generates fuzzing harnesses for Rust libraries, using static analysis augmented with LLMs, for the purpose of finding memory-safety bugs. In an evaluation over 63 real-world crates, deepSURF covers 87.3\% of URAPIs and rediscovers 30 of 56 known bugs, plus 12 new memory-safety bugs, outperforming existing Rust fuzzers.

\subsection{Scientific Contributions}
\begin{itemize}
\item Provides a Valuable Step Forward in an Established Field
\end{itemize}

\subsection{Reasons for Acceptance}
\begin{enumerate}
\item The paper provides a valuable step forward in an established field. Compared to prior Rust fuzzing tools, deepSURF fuzzes more of the target library (and finds more memory-safety bugs) by generating closures for APIs that require them, and invoking related APIs to create different states when invoking target URAPIs; these advances are made possible by a creative combination of static analysis and LLMs. The paper’s empirical evaluation is well done.
\end{enumerate}

\subsection{Noteworthy Concerns}
\begin{enumerate}
\item The paper’s evaluation presents a strong quantitative result, but could do more to readily show the effectiveness of deepSURF by qualitatively analyzing the generated harnesses and systematically comparing them to actual usages of the targeted APIs, to assess realism.
\end{enumerate}

\section{Response to the Meta-Review}
A preliminary qualitative analysis indicates that \mbox{deepSURF}'s generated harnesses reflect realistic usage patterns and adhere to the documented APIs and semantics of the evaluated crates. For example, for crates implementing data structures such as \texttt{toodee}, \texttt{smallvec}, \texttt{slice\_deque}, \texttt{simple-slab}, and \texttt{stack\_dst}, deepSURF produces harnesses that naturally follow common user workflows: constructing the data structure, populating it with insertion operations, and subsequently performing stateful transformations such as removals, cloning, swapping, and filtering.
However, conducting a systematic qualitative assessment of the generated harnesses (whether through manual review or automated analysis) would require substantial additional effort. We will explore this as part of future work.
\end{document}